\documentclass{mn2e}
\usepackage{epsfig}

\usepackage{graphicx}
\newif\ifAMStwofonts
\AMStwofontstrue

\title[Strong biases in infrared-selected gravitational lenses]
{Strong biases in infrared-selected gravitational lenses}
\author[Serjeant]
{Stephen Serjeant$^1$\\
$^1$Department of Physical Sciences, The Open University, Milton
Keynes, MK7 6AA, UK
} \date{Received 2011} \pubyear{2011}

\begin{document}

\input BoxedEPS.tex
\SetEPSFDirectory{./}

\SetRokickiEPSFSpecial
\HideDisplacementBoxes

\label{firstpage}

\maketitle

\newcommand{\bfreferee}[0]{}

\begin{abstract}
  Bright submillimetre-selected galaxies have been found to be a rich
  source of strong gravitational lenses. However, strong gravitational
  lensing of extended sources leads inevitably to differential
  magnification.  In this paper I quantify the effect of differential
  magnification on simulated far-infrared and submillimetre surveys of
  strong gravitational lenses, using a foreground population of
  Navarro-Frenk-White plus de Vaucouleurs' density profiles, with a
  model source resembling the Cosmic Eyelash and QSO J1148$+$5251.
  Some emission line diagnostics are surprisingly unaffected by
  differential magnification effects: for example, the bolometric
  fractions of [C{\sc ii}] $158\,{\rm \mu}$m and CO$(J=1-0)$, often used to
  infer densities and ionisation parameters, have typical differential
  magnification effects that are smaller than the measurement
  errors. However, the CO ladder itself is significantly
  affected. Far-infrared lensed galaxy surveys (e.g. at $60\,{\rm \mu}$m)
  strongly select for high-redshift galaxies with caustics close to
  active galactic nuclei (AGN), boosting the apparent bolometric
  contribution of AGN. The lens configuration of IRAS F10214$+$4724 is
  naturally explained in this context.  Conversely, {\bfreferee
    submillimetre/millimetre-wave surveys (e.g. $500-1400\,{\rm \mu}$m)}
  strongly select for caustics close to knots of star formation
  boosting the latter's bolometric fraction. In general, estimates of
  bolometric fractions from spectral energy distributions of strongly
  lensed infrared galaxies are so unreliable as to be useless, unless
  a lens mass model is available to correct for differential
  magnification.
\end{abstract}


\section{Introduction}\label{sec:introduction}
Strong gravitational lensing has been widely exploited for studing
populations that would otherwise be too faint for detailed study
(e.g. Smail et al. 1997, Knudsen et al. 2008), as well as being one of
the very few probes capable of mapping the dark matter distributions
in the foreground systems (e.g. Gavazzi et al. 2007). Gravitational
lenses were originally very difficult to discover: for example, the
Cosmic Lens All-Sky Survey (CLASS) observed 11\,685 flat-spectrum radio
sources, finding 16 gravitational lenses (Myers et al. 2003). More
rapid lens discovery has come from exploiting the large public Sloan
Digital Sky Survey (SDSS) database. The Sloan Lens ACS Survey (SLACS)
selected lens candidates for {\it HST} snapshot follow-up from the
presence of nebular emission lines at a higher redshift than that of
the SDSS galaxy (e.g. Bolton et al. 2006), resulting in nearly 100
published lenses (Auger et al. 2009). The SLACS lens redshifts are
$z<0.5$ by virtue of their selection criteria, which limits the study
of the evolving mass profiles of the lenses, while the background
galaxies are by necessity unobscured blue star-forming systems with
sufficiently dust-free lines of sight through the foreground lens.
Many ongoing and future applications of gravitational lensing are also
limited by sample size (e.g. Treu 2010), and so there are strong
drivers both to finding lenses at higher redshifts and to finding
lensed systems in greater numbers.

{\bfreferee Several authors} predicted that the steep bright-end slope
of submillimetre source counts, together with the high redshifts of
submillimetre-selected galaxies {\bfreferee (e.g. Chapman et al. 2005,
  Eales et al. 2010)} would lead to a strong gravitational
magnification bias {\bfreferee (e.g. Blain 1996, Perrotta et al. 2002,
  Negrello et al. 2007).}  Shallow, wide-area submillimetre surveys
could therefore be exploited as an extremely efficient means of
selecting strong lens events, once the contaminant populations of
nearby galaxies and blazars are removed. The millimetre-wave source
counts from the South Pole Telescope {\bfreferee (SPT)} were
consistent with this prediction (Vieira et al. 2010), and the
prediction was spectacularly confirmed by Negrello et al. (2010), who
used an intensive and rapid follow-up of the first data from the {\it
  Herschel} Astrophysical Terahertz Large Area Survey ({\it H}-ATLAS,
Eales et al. 2010) to find that all five of the first lensed galaxy
candidates are strong gravitational lenses. The selection efficiency
for strong lenses in bright submillimetre surveys approaches an
astonishing $\simeq 100\%$. The lens redshift distribution is expected
to peak at $z\stackrel{>}{_\sim}1$, with lensing systems selected
irrespective of the obscuration in the foreground system.  {\bfreferee
  The advent of wide-field mapping capabilities at submillimetre and
  millimetre wavelengths from e.g. {\it Herschel} Spectral and
  Photometric Imaging Receiver (SPIRE), Submillimetre Common-User
  Bolometer Array-2 (\mbox{SCUBA-2}) and SPT allow the discovery of
  many hundreds of strong lenses (e.g. Vieira et al. 2010, Negrello et
  al. 2010, Cooray et al. 2010, Gonz\'{a}lez-Nuevo et al. 2012).}
This immediately also affords many opportunties for follow-up imaging
and spectroscopic diagnostics of the lensed galaxies (e.g.  Lupu et
al. 2010, Cooray et al. 2011, Cox et al. 2011, Frayer et al. 2011,
Gavazzi et al. 2011, Hopwood et al. 2011, Omont et al. 2011, Riechers
et al. 2011, Scott et al. 2011, Valtchanov et al. 2011).

Gravitational lensing is a purely geometrical effect depending on the
paths of null geodesics, and is therefore wavelength-independent.
Nevertheless, different lines of sight have different magnifications
in every astrophysical lens. Therefore if the background source is
extended with intrinsic colour gradients, the lensed source may have
colours 
%not representative of 
that are different to 
the unlensed case. This differential
magnification could significantly affect broad-band photometry and
emission line diagnostics. Many strong gravitational lenses do not
have a well-constrained foreground mass model, so the level and nature
of the differential magnification problem in those sources is unknown.
Often, a largely unjustified assumption is made, explicity or
implicitly, that differential magnification effects can be neglected,
{\bfreferee despite the fact that it has long been clear that differential
  magnification effects are present and must be corrected for
  (e.g. Blain 1999).}

This paper presents a statistical assessment of the level of the
differential magnification problem for a wide range of photometric and
spectroscopic diagnostics of the physical conditions in
gravitationally lensed infrared-luminous galaxies. Section \ref{sec:method}
describes the simulation methodology, including the foreground lenses
and the adopted background source. Section \ref{sec:results} gives the
results of the simulations, demonstrating where it is and is not
appropriate to assume a naive demagnification without differential
magnification effects. Section \ref{sec:discussion} discusses the
implications of these results for ongoing follow-up studies of
infrared-selected strong gravitational lenses,
{\bfreferee and the conclusions are summarised in Section
  \ref{sec:conclusions}.}
This paper assumes a
concordance cosmology with a Hubble constant of
$H_0=72$\,km\,s$^{-1}$\,Mpc$^{-1}=100h$\,km\,s$^{-1}$\,Mpc$^{-1}$, a
matter density of $\Omega_{\rm M}=0.3$ and a cosmological constant
effective energy density of $\Omega_\Lambda=0.7$.

\section{Methodology}\label{sec:method}
\subsection{Simulation details}
Differential magnification effects were simulated for a given source at
a fixed redshift, seen through a constant comoving population of
lenses. 
The physical parameters of the lens and the source in this
case are given in the following subsections. The source plane
magnification map, summing over all images $\Sigma_i|\mu_i(x,y)|$, was calculated using the
{\sc gravlens} 
software (Keeton 2001). The simulation resolution was
$0.01$ arcsec at this stage. With the 
source redshift
fixed, the differential 
{\bfreferee magnification}
of the various
source components was used to calculate observed spectral energy distributions
(SEDs). A selection at a particular wavelength was then
imposed: I first assumed that the background source will be considered a
strong gravitational lens {\it at that observed wavelength} when the total
magnification over all source components exceeds a threshold
{\bfreferee magnification} 
$\mu\ge 10$. This was then compared to the case of more
moderate magnifications of $2<\mu<5$ at that wavelength. Most of the results 
presented in this paper are for a $z=2$ background source, but it was
found that the results depend only weakly on the background source
redshift. 
{\bfreferee Differential magnification effects depend on the {\it relative}
  angular sizes of the magnification map and the background source. As
  shown in Fig.\,\ref{fig:thetae}, angular diameter distance varies
  only weakly with redshift at $z\simeq1-3$ (see e.g. Serjeant 2010),
  while the observed critical radii for singular isothermal spheres
  depend only very weakly on the source and lens redshifts over most
  of their range (unlike the point mass case). It should therefore not
  be too surprising that there is no strong dependence on the source
  redshift in these simulations, notwithstanding changes in rest-frame
  wavelengths. Section \ref{sec:various_zsource} presents results for
  a variety of source redshifts. Most of the results in this paper are
  based on lenses selected at observed wavelengths of either
  $60\,{\rm \mu}$m or $500\,{\rm \mu}$m, representing selection with e.g. {\it
    IRAS} and the {\it Herschel} SPIRE instrument. The $500\,{\rm \mu}$m
  results for a source at $z=2$ are very similar in practice to
  simulations at $850\,{\rm \mu}$m or $1.4$\,mm, representing lens
  selection with e.g. SCUBA-2 or the SPT; this is because all three
  wavelengths lie well within the redshifted Rayleigh-Jeans
  tail. Section \ref{sec:various_zsource} discusses the dependences on
  observed wavelength in more detail.
}

A higher resolution ($0.001$ arcsec) simulation was then run for the
specific case of a lens geometry based on that of
IRAS\,FSC\,10214+4724, with a single lens at a redshift of $z=0.9$ and
a source at $z=2.286$. Finally, calculations were made of flux- or
luminosity-limited samples in the context of a particular source count
model.

{\bfreferee 
\begin{figure}
\centering
\ForceWidth{7in}
\hSlide{-4cm}
%%\vspace*{-1cm}
\BoxedEPSF{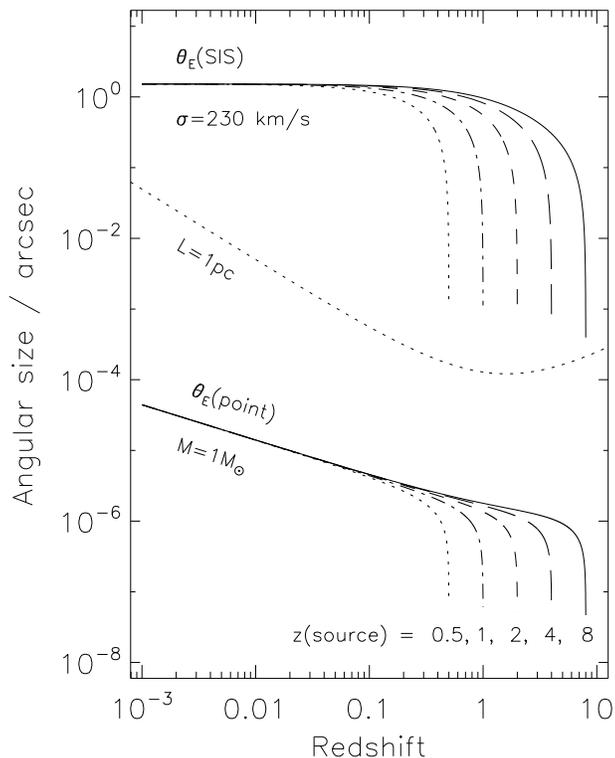}
%%\vspace*{-1cm}
\caption{\label{fig:thetae} The Einstein radius in arcseconds against
  lens redshift for a singular isothermal sphere with velocity
  dispersion equal to the SLACS mean of $230\,$km\,s$^{-1}$ (upper
  curves, Bolton et al. 2008), and for comparision a one solar mass
  point lens (lower curves), at source redshifts of $0.5$, $1$, $2$,
  $4$ and $8$ (dotted, dash-dot, dashed, long dashed and full lines
  respectively). Note that the Einstein radius for the singular
  isothermal sphere has a much weaker dependence on the lens redshift
  than the point mass case, until the lens approaches the source.  As
  a further comparison, the dotted line shows the observed angular
  size of a fiducial one parsec proper length. Note that over
  $z\simeq0.5-3$, the observed angular sizes are relatively weakly
  dependent on redshift. The weak dependence on redshift of both the
  source sizes and lens critical radii imply the results in this paper
  will be only weakly dependent on the background source redshift,
  notwithstanding differences in rest-frame wavelengths.}
\end{figure}
}

\subsection{The foreground lenses}\label{sec:lens_model}
The foreground lensing galaxy or galaxies were simulated following
SLACS (Bolton et al. 2006). The assumed density 
profile was the sum of a Navarro-Frenk-White (NFW) profile and a 
de Vaucouleurs' profile, which has an azimuthally-averaged surface
density similar to a singular isothermal ellipsoid (e.g. Gavazzi et
al. 2007). 
{\bfreferee Spiral galaxy discs present only $10-20\%$ of the lensing
  cross-section of elliptical galaxies (e.g. Keeton \& Kochanek 1998,
  M\"{o}ller et al. 2007) so spiral discs are neglected in this
  analysis.  Also neglected for simplicity is the lensing by galaxy
  groups and clusters, in which it may be rarer to find
  subarcsecond-scale strong magnification gradients intercepting the
  background sources. Nonetheless, the existence of giant optical arcs
  implies that galaxy clusters can sometimes induce strong
  magnifications (e.g. Bayliss et al. 2011, Wuyts et al. 2012). In
  these situations, one might expect differential magnification to be
  relevant.}

The NFW  
{\bfreferee profiles of the elliptical galaxy haloes in this paper}
had a virial mass equal to the mean SLACS lens
mass, $M_{\rm vir}=14h^{-1}\times10^{12}M_\odot$, and a concentration parameter $c$ 
given by 
\begin{equation}\label{eqn:concentration}
c = \frac{9}{1+z}\left (\frac{M_{\rm vir}}{8.12 h^{-1}M_\odot}\right
)^{-0.14}
\end{equation}
(Hoekstra et al. 2005, correcting a typographical error in the $h$
dependence, e.g. Bullock et al. 2001, Gavazzi et al. 2007). 
The virial radius was set to $R_{\rm vir}=466/h$\,kpc, consistent with
\begin{equation}
M_{\rm vir}=\frac{4{\rm \pi}\Delta}{3}\rho_{\rm c}R_{\rm vir}^3
\end{equation}
where $\rho_{\rm c}$ is the critical density and $\Delta=119$. 
For the de Vaucouleurs' profile, the
effective radius was set to the mean of the SLACS lenses (Gavazzi et al. 2007), 
$5.3h^{-1}$\,kpc. 

For the case of a single lens, the configuration was modelled on the
lensed hyperluminous galaxy IRAS F10214$+$4724 (Rowan-Robinson et
al. 1991, Broadhurst \& Leh\'ar 1995, Graham \&
Liu 1995, Serjeant et al. 1995, 
Eisenhardt et al. 1996), with a background source at a
redshift of $z=2.286$ (e.g. Brown \& Vanden Bout 1991) and a lens at
$z=0.9$ (e.g. Serjeant et al. 1995, 1998). 
The ellipticity of the lens, defined as one minus the axis ratio, 
was set to $e=0.21$, equal to the median value for SLACS
lenses (Bolton et al. 2008). No external shear was included as only one 
angle dependence is necessary to reproduced the SLACS lens
morphologies. Under these assumptions, the 
critical (Einstein) radius is $1.18$ arcsec, close to the median
value of $1.17$ for the SLACS singular isothermal ellipsoid models
{\bfreferee (table 5 of Bolton et al. 2008).}

For the case of a population of lenses, a constant comoving
density of these SLACS-like lenses was assumed, whose mass profiles do not
evolve except in the variation of concentration parameter (equation
\ref{eqn:concentration}). No
assumptions were made about the comoving density of background sources;
instead, the differential lensing statistics were found as a function
of source redshift. 

{\bfreferee 
\begin{figure}
\centering
\ForceWidth{3.5in}
%%\hSlide{-1.8cm}
%%\vspace*{-1cm}
\BoxedEPSF{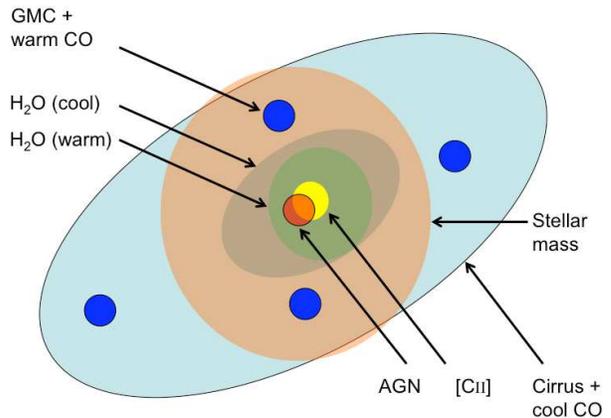}
%%\vspace*{-1cm}
\caption{\label{fig:cartoon}
A sketch, not to scale, of the emission regions in the model
background source described in Section \ref{sec:source_model}. The
GMCs are embedded in larger warm CO regions.  
}
\end{figure}
}

\subsection{The background source}\label{sec:source_model}
\subsubsection{SMG model}
I modelled a submillimetre-selected galaxy (SMG) -like extended
background source with a structure resembling that of the Cosmic
Eyelash (Swinbank et al. 2010) and the $z=6.42$ quasar QSO
J1148$+$5251 (e.g. Walter et al. 2009).  {\bfreferee The objective is
  to calculate the differential magnification effects on common
  emission-line diagnostics (CO, [C{\sc ii}], H$_2$O, etc.) and
  bolometric fractions (AGN, star formation, cirrus) so the model
  background galaxy has spatially localised regions and associated
  SEDs for each of these emission components.  Fig.\,\ref{fig:cartoon}
  shows a sketch (not to scale) of the components of this simulated
  galaxy.}  The source has the following features:
\begin{enumerate}
\item An AGN with a bolometric fraction of $0.3$, with an
SED from Nenkova et al. (2008a,b), modelled as a circular region with
$0.1$\,kpc radius. 
The results are not strongly dependent on the AGN torus model chosen,
as long as its SED peaks in the rest-frame mid-infrared. 
The selected model has an AGN and dust torus, 
incident AGN spectum AA01, 
an angular distribution width $\sigma=30^\circ$, 
a Gaussian angular dust distribution with an equivalent width of
$45^\circ$, 
a radial torus thickness of $100$ sublimation radii, 
an average of $14$ clumps along a radial equatorial ray, 
a cloud distribution with a radial dependence of $r^{-2.0}$, and a
visual optical depth of individual clumps of $\tau_{\rm V}=100$. 
The torus is simulated as viewed from an inclination angle of
$50^\circ$. 
\item An extended cirrus region with a $2.5\,$kpc major axis, an
  ellipticity of $e=0.4$, a bolometric fraction of $0.2$, and a cirrus
  SED from Efstathiou, Rowan-Robinson \& Siebenmorgen (2000). 
\item Four extreme giant molecular cloud (GMC) systems,
as seen in the Cosmic Eyelash, with a total
bolometric fraction of $0.5$. The GMCs (assumed identical) 
are placed randomly into the
cirrus region, and have a starburst SED taken from the models of Efstathiou,
Rowan-Robinson \& Siebenmorgen (2000), with an age of $1.7$\,Myr and
an optical depth of $\tau_{\rm V}=100$. They are modelled as $50$\,pc
radius circular clouds. 
\item A circular old stellar population with a 2\,kpc half-light
  radius and a de Vaucouleurs' profile. The 2\,Gyr-old elliptical SED
  from the {\it Spitzer} Wide-Area Extragalactic Survey (SWIRE)
  library (Polletta et al. 2007) is adopted for this component, though
  for this study it is only necessary that some relatively AGN-free
  observational proxy is available (e.g. {\it Spitzer} $3.6\,{\rm \mu}$m
  fluxes) and that it is possible to subtract the foreground lens
  (e.g. Hopwood et al. 2011).
\item A circular [C{\sc ii}] emitting region offset $0.6\,$kpc from the AGN
with radius $0.75\,$kpc, following the precedent of SDSS QSO J1148+5251.
Similarly, the lensed star-forming galaxy MIPS J142824.0+352619 has
been argued to contain a [C{\sc ii}]-emitting region extended over kpc
scales (Hailey-Dunsheath et al. 2010). 
\item HCN and HNC emission regions are modelled as two point sources
  separated by $80$\,pc (see Section \ref{sec:hcn}). 
\item Two H$_2$O emission regions: a circular ``warm'' region with a
  radius of $120$\,pc, and an elliptical ``cool'' region with a major
  axis of $1$\,kpc and an ellipticity and position angle matching the
  cirrus component. This follows the precedent of Mrk\,231 (van der
  Werf et al. 2010, Gonz\'alez-Alfonso et al. 2010). 
\item A cool molecular gas component tracing the cirrus. The CO
  emission spectrum was calculated using the RADEX software (van der
  Tak et al. 2007) using the Large Velocity Gradient (LVG) approximation, 
  assuming it is comprised of clumps in a background
  temperature of $8.97$\,K (i.e. $2.73(1+z)$\,K at $z=2.286$), with a
  kinetic temperature of $20$\,K, an H$_2$ density of
  $100$\,cm$^{-3}$ %% see eg http://arxiv.org/abs/astro-ph/0507587
  and a CO column density of $10^{14}$\,cm$^{-2}$ for each cirrus
  cloud.
\item A warm molecular gas knot with $0.4$\,kpc radius centred on each
  GMC. Following the precedent of SDSS QSO J1148+5251, these warm
  components in total contribute $1/3$ of the total CO$(J=3-2)$
  flux. The CO emission of these knots was calculated using the RADEX (van
  der Tak et al. 2007) LVG model assuming they are comprised of clumps in a
  background temperature of $8.97$\,K (i.e. $2.73(1+z)$\,K at $z=2.286$),
  with a kinetic temperature of $45$\,K
  an H$_2$ density of $10^4$\,cm$^{-3}$ and a CO column density of
  $10^{18}$\,cm$^{-2}$. 
  {\bfreferee There is already good evidence that the higher $J$ transitions of
    CO in submillimetre-selected galaxies are much more spatially localised
    than CO$(J=1-0)$ (e.g. Ivison et al. 2011).  }
\end{enumerate}
This model source will be referred to as the SMG model. In the
simulations with $0.01$ arcsec resolution, the maximum monochromatic
magnification of a $500\,{\rm \mu}$m-selected lens at $z=2$ for a constant
comoving density of lenses is $\mu=33$. Restricting this to
configurations with magnifications $\mu>10$, the median (mean)
magnification is $11.5$ ($12.0$). At an observed wavelength of
$60\,{\rm \mu}$m the flux is dominated by much smaller scale features such
as the AGN, so the corresponding numbers are higher: the median (mean)
magnification of $\mu>10$ sources is $18.5$ ($21.9$), while the
maximum monochromatic magnification is $840$. The configurations that
approach this (possibly unphysical) maximum are nonetheless extremely
rare, and the results in this paper are unaffected by excluding all
configurations with magnifications of e.g. $\mu>100$. The higher
angular resolution simulation discussed in Section \ref{sec:f10214}
has a more moderate maximum magnification.

\subsubsection{Filling factors}

The filling factor of emission line gas in SMGs is far higher than in
local galaxies. For example, Carilli et al. 2010 derived a filling
factor of 0.5 in the low excitation gas component of their $z=4.05$
SMGs, and 0.13 in its higher excitation gas. These values are an order
of magnitude or more larger than seen in local galaxies (e.g. McKee \&
Ostriker 2007). In effect, the clumping of the emission line gas in
high-redshift star-forming galaxies creates a sparse sampling of the
morphological distributions defined above. Since the surface filling
factor is high, there are unlikely to be problems caused by small
number statistics of clumps.  The objective is to simulate a
high-redshift galaxy, and therefore in 
the SMG model 
the emission line gas is treated as spatially
continuous.  Any further clumping beyond that already specified is
neglected.

\begin{figure*}
\centering
\ForceWidth{4.47in}
\hSlide{-3.1cm}
%\BoxedEPSF{lines1.ps}
\BoxedEPSF{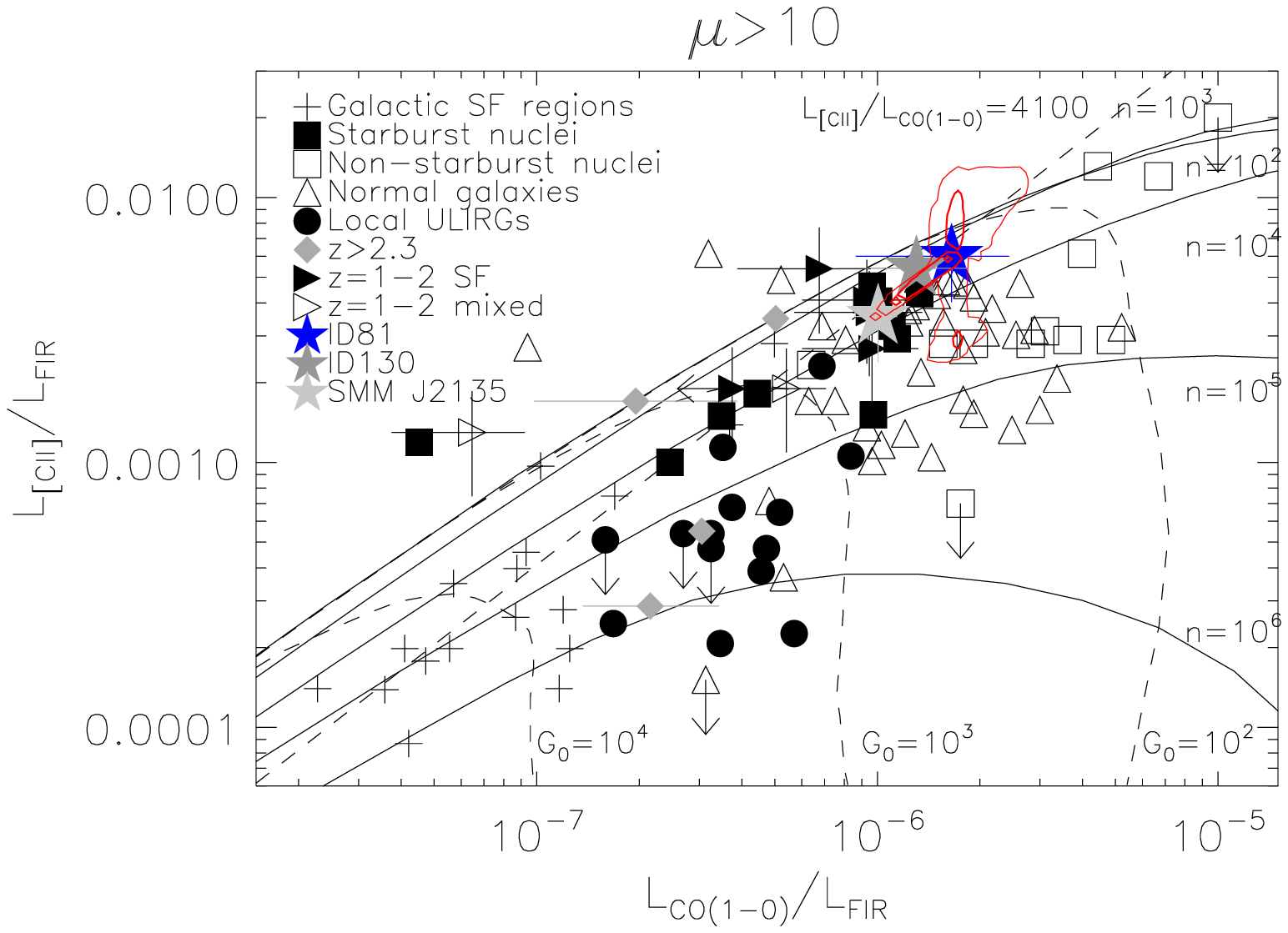}
\ForceWidth{3.15in}
\hSlide{5.1cm}
\vSlide{-7.38cm}
%\BoxedEPSF{lines2_zs=2.00000_sparse=1_magthresh=10.0000_magmax=1.0000000e+10.ps}
\BoxedEPSF{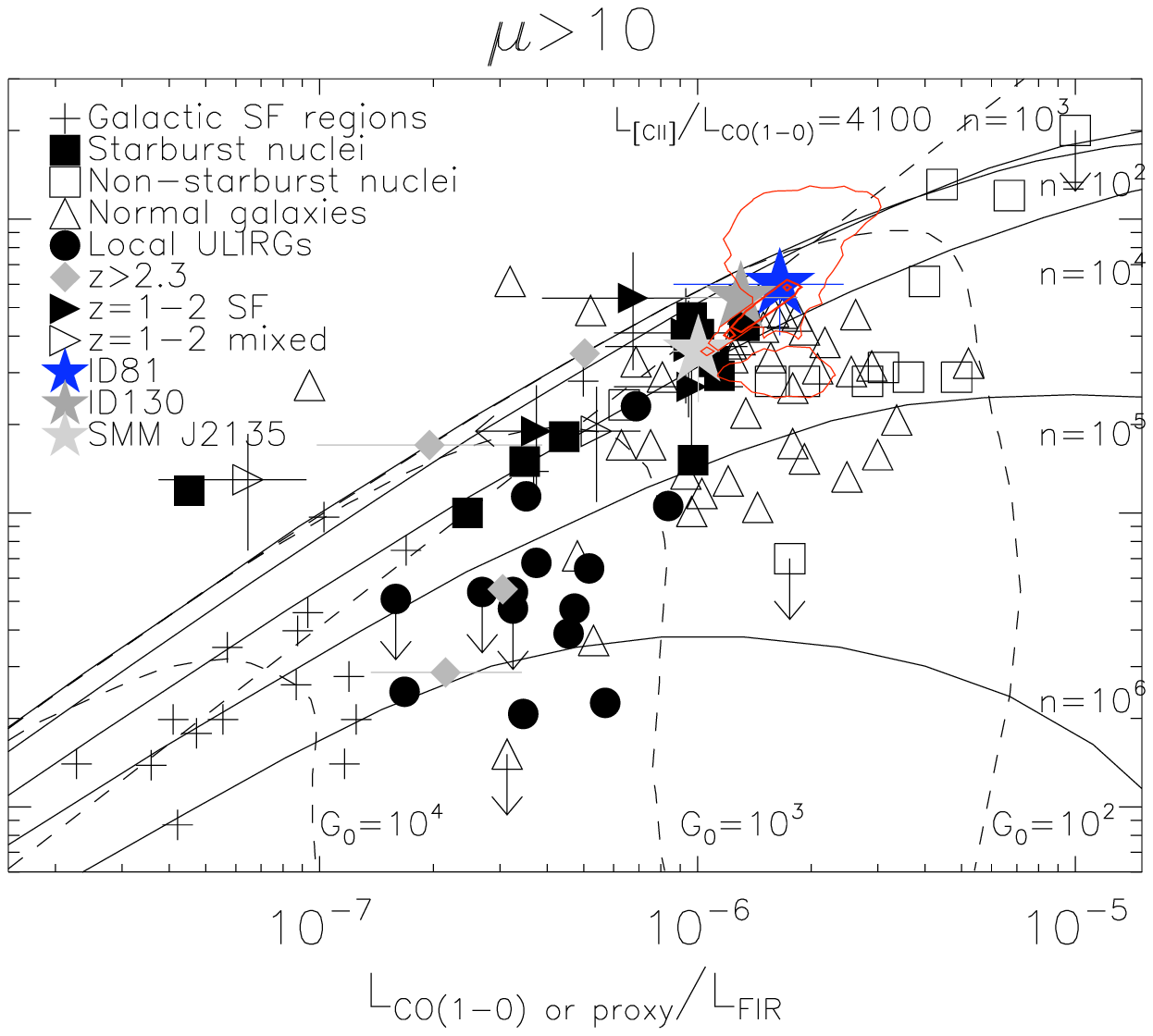}
\vspace*{-8cm}
\ForceWidth{4.47in}
\hSlide{-3.1cm}
\BoxedEPSF{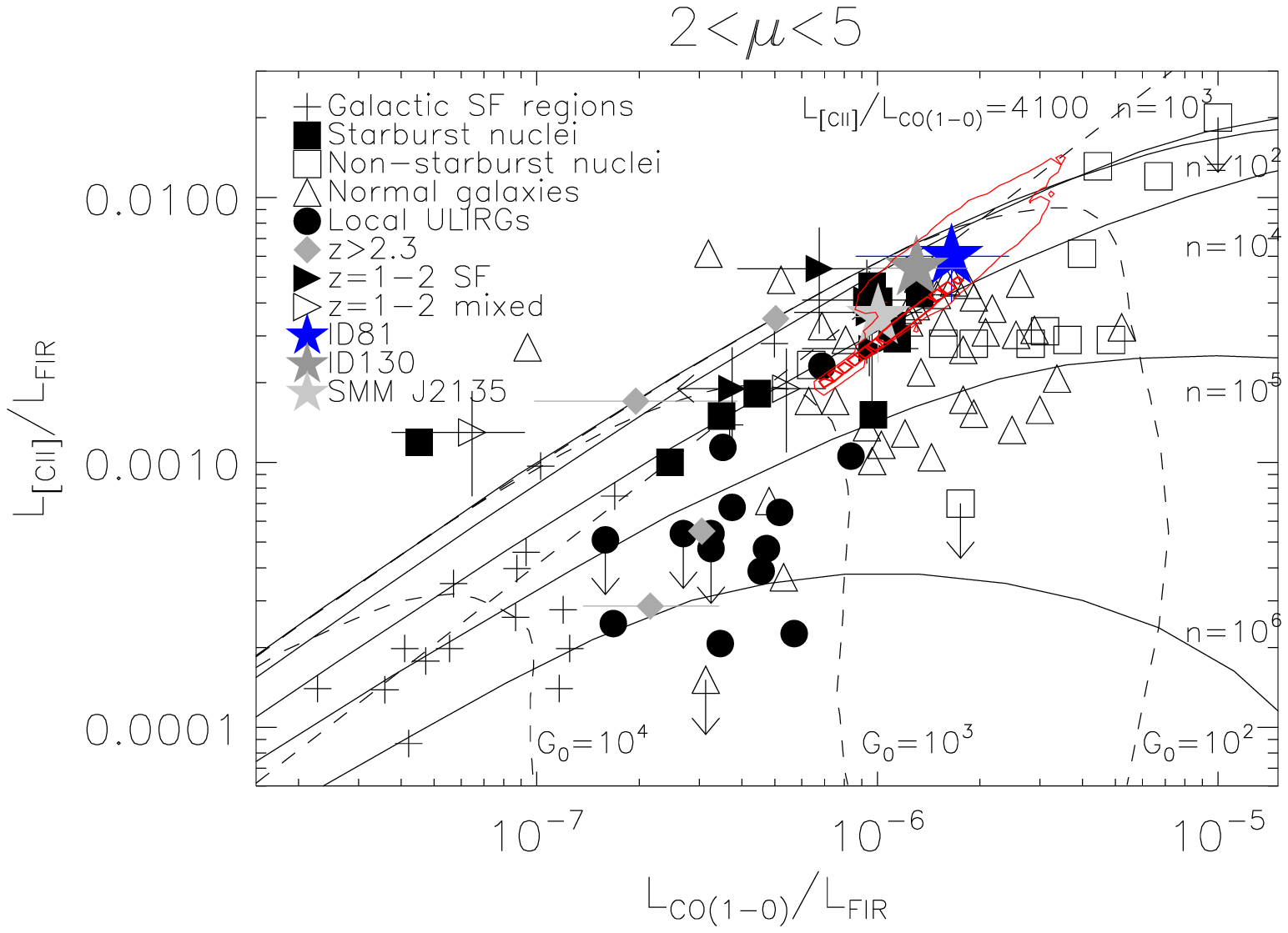}
\ForceWidth{3.15in}
\hSlide{5.06cm}
\vSlide{-7.38cm}
%\BoxedEPSF{lines2_zs=2.00000_sparse=50_magthresh=2.00000_magmax=5.00000.ps}
\BoxedEPSF{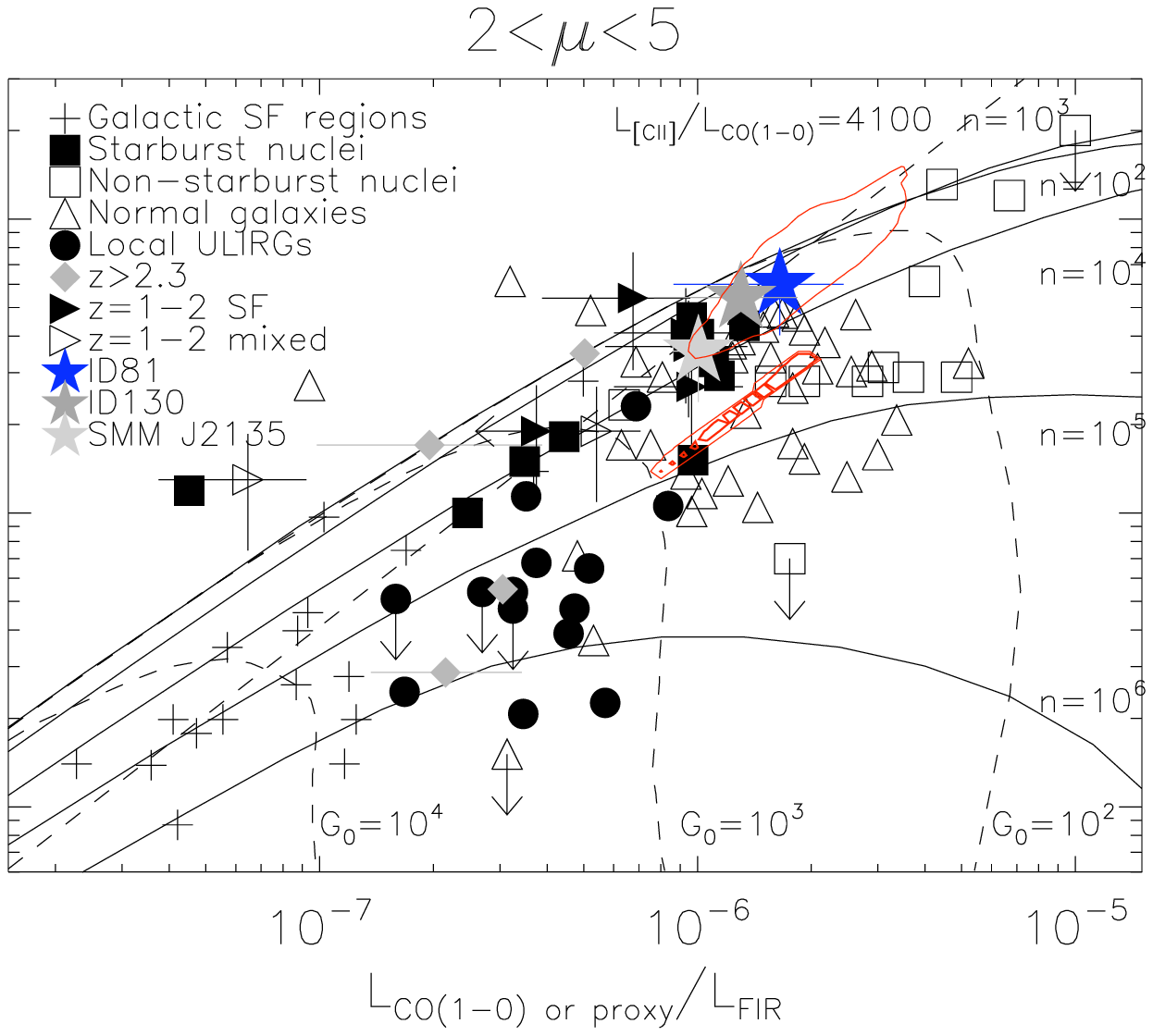}
\vspace*{-7cm}
\caption{\label{fig:cii}
Simulated emission line diagnostic diagram for a $500\,{\rm \mu}$m-selected
gravitational lens with a source at $z=2$; very similar results were
found for sources at other redshifts. The underlying value of the
line/FIR 
ratios is taken to be identical to the H-ATLAS galaxy ID81
(Valtchanov et al. 2011), while the differentially magnified simulated
data points are shown as 
{\bfreferee red contours 
near the ID81 data
point (contours at $68\%$, $95\%$, $99\%$ and $99.9\%$).  
}
The observed data are taken from the
compilation in Stacey et al. 2010 and Valtchanov et al. 2011. Also
shown are densities and interstellar far-ultraviolet flux predictions from Stacey
et al. 2010; densities are in units of cm$^{-3}$ and the ionising flux
$F_{\rm FUV}$
is parameterised as $G_0=F_{\rm
  FUV}/(1.6\times10^{-6}$\,W\,m$^{-2})$. 
The left-hand panels show the use of the CO$(J=1-0)$
line, while the right-hand panels show the use of CO$(J=4-3)$ as a proxy
for CO$(J=1-0)$. The lower panels show results for moderate
magnifications of $2<\mu<5$, while upper panels show high
magnification systems of $\mu>10$. 
{\bfreferee 
In all four cases, nearly all the contours follow a track that is
almost parallel to a constant density 
diagnostic curve. The vertical positioning of this track was
found to depend on the source structure.
}
}
\end{figure*}

\begin{figure*}
\centering
%\ForceWidth{4.5in}
%\vspace*{-1cm}
%\BoxedEPSF{co_sled1.ps}
\ForceWidth{3.4in}
\hSlide{0.1cm}
\BoxedEPSF{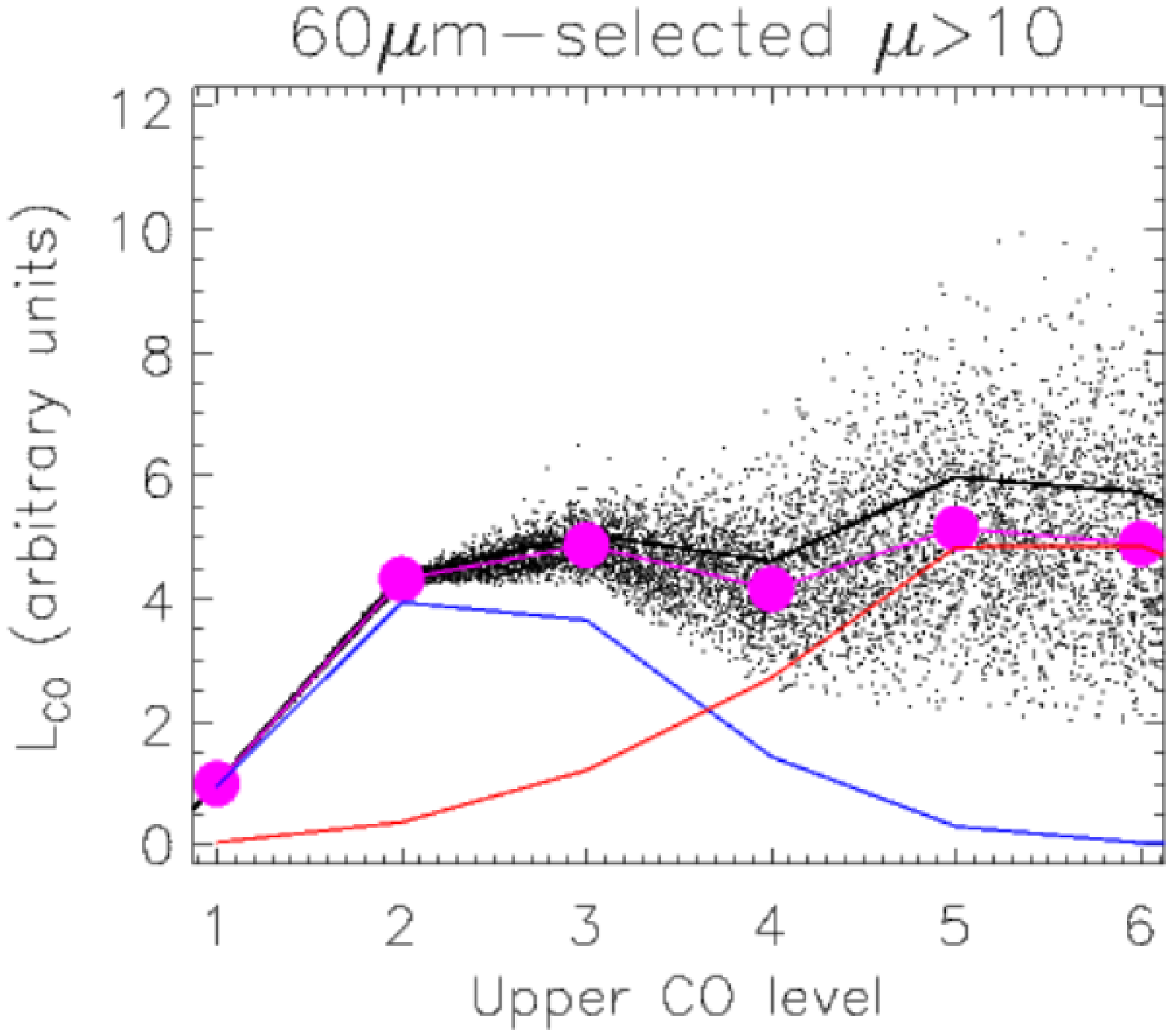}
\ForceWidth{3.4in}
\hSlide{-0.2cm}
\BoxedEPSF{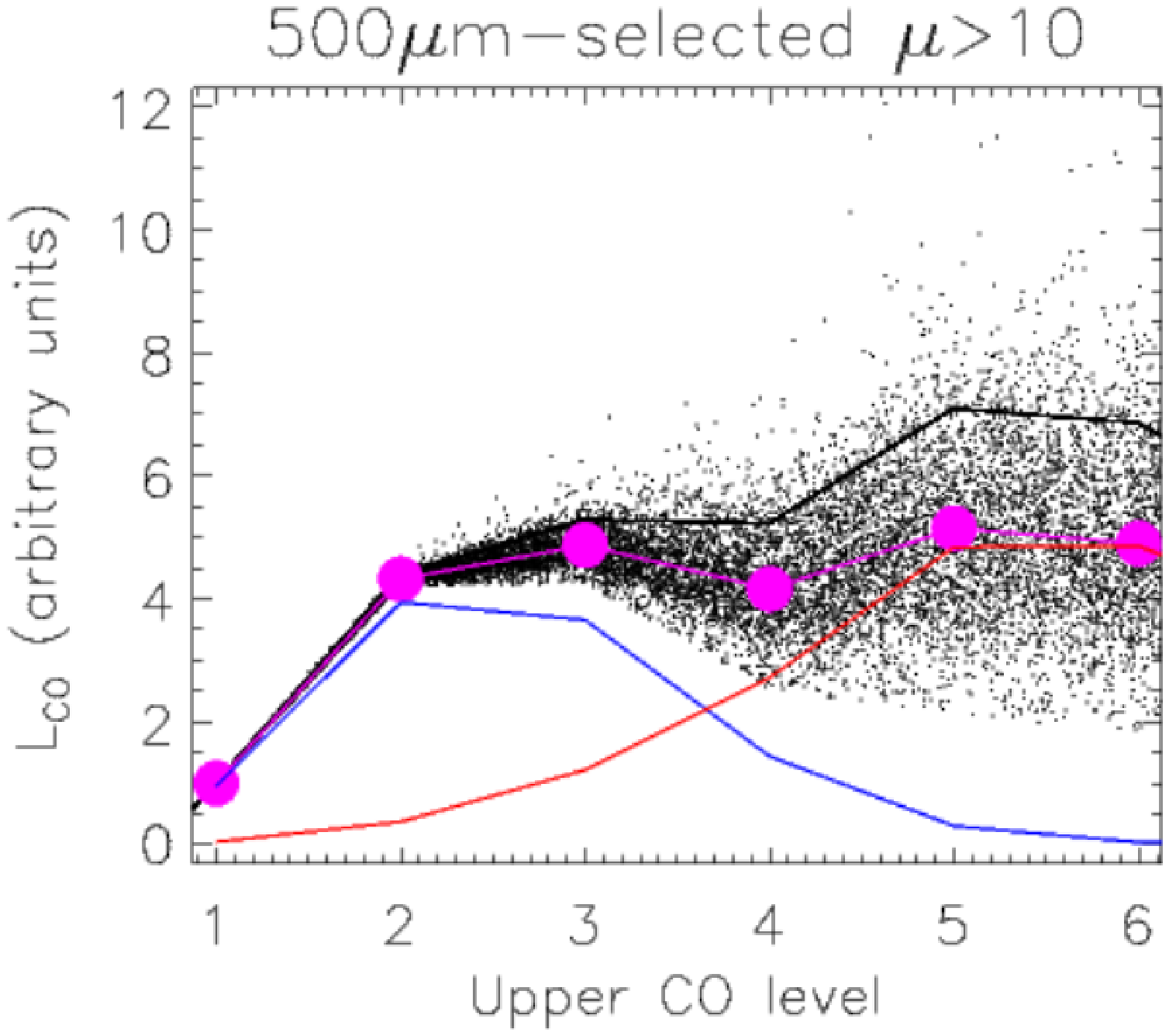}
%\vspace*{-1cm}
\ForceWidth{3.4in}
\BoxedEPSF{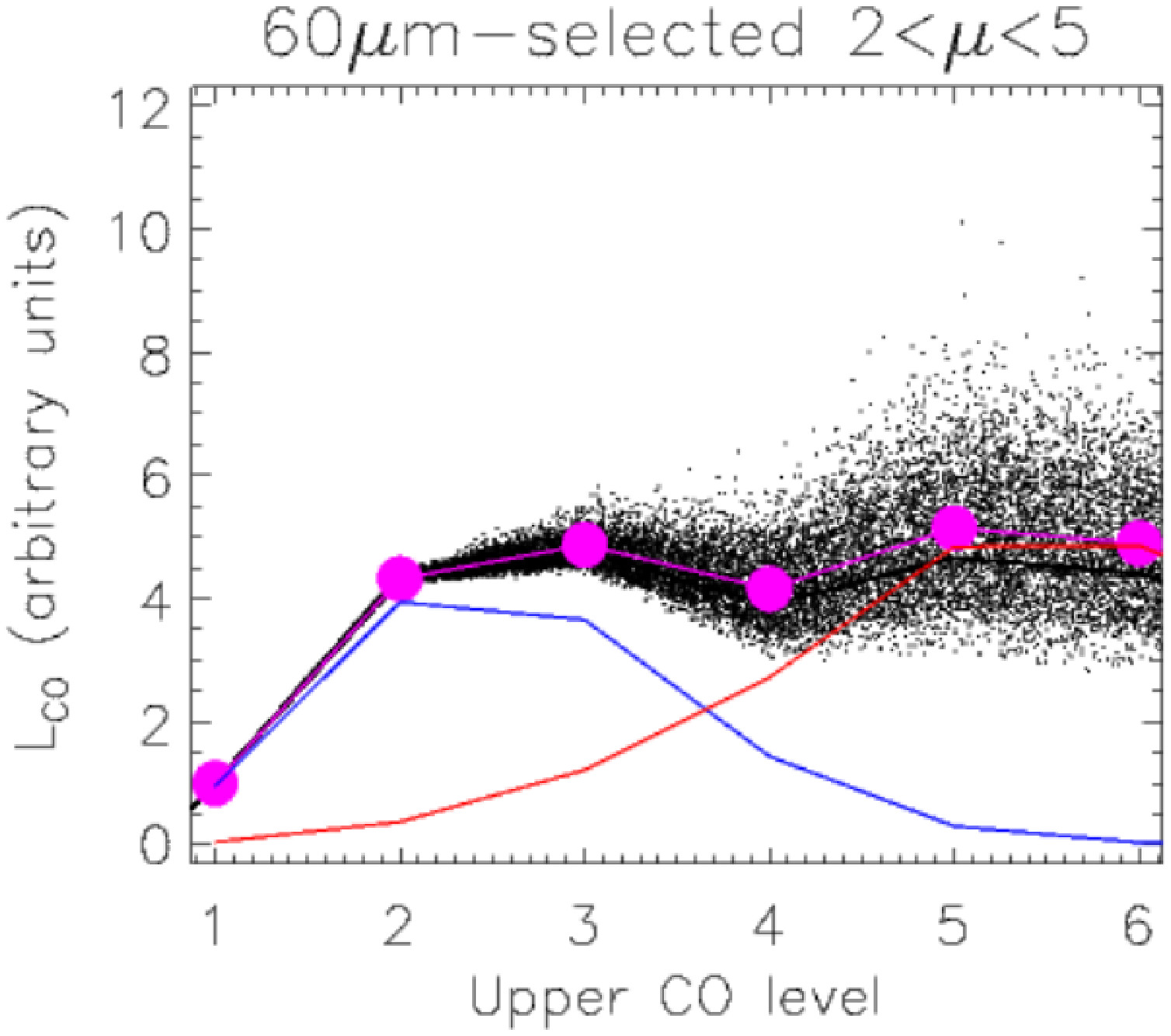}
\ForceWidth{3.4in}
\BoxedEPSF{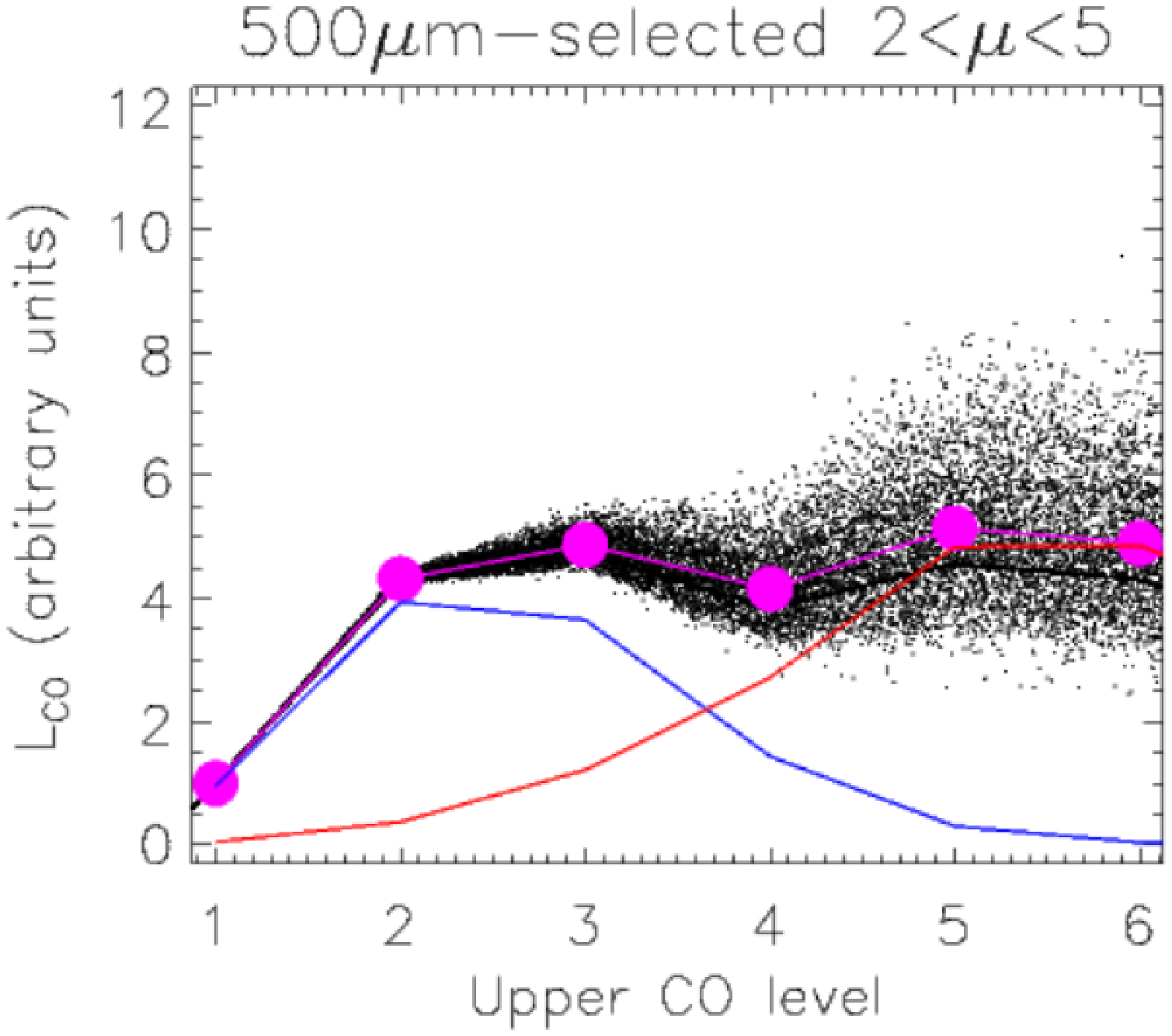}
%\vspace*{-1cm}
\caption{\label{fig:co_sled} Simulated CO Spectral Line Energy
  Distribution (SLED) for a $z=2$ gravitational lens, selected at an 
  observed wavelength of $60\,{\rm \mu}$m (left) and $500\,{\rm \mu}$m (right), 
  using the SMG model as the source. Very similar results were
  obtained for other source redshifts. The upper panels show the SLEDs
  for high-magnification lenses with $\mu>10$, while the lower panels
  show the SLEDs for more moderate magnifications of $2<\mu<5$; only a
  random sparse sample of 2 (10) percent of the lens configurations are shown
  in the moderate (strong) magnification case, for clarity. 
  The unlensed cool component is
  shown as the \emph{blue line}, while the unlensed warm component is 
  a \emph{red line}. The \emph{magenta line and 
  large symbols} show the sum of the
  unlensed compoents. The total lensed CO SLEDs are shown as \emph{dotted
  lines}, with the dots spaced randomly along each line to avoid
  overlapping points. 
{\bfreferee 
In each case, one particularly favoured configuration 
  leads to multiple superimposed dotted lines (i.e. the black line).}
All SLEDs have been normalised to the
  CO$(J=1-0)$ point, to remove the the overall effect of magnification
  and isolate the effects of differential magnification.  }
\end{figure*}

\section{Results}\label{sec:results}
\subsection{Emission line diagnostics: [C{\sc ii}] and [O{\sc i}]}
Fig.~\ref{fig:cii} shows a simulated emission line diagnostic diagram
for a $500\,{\rm \mu}$m-selected SMG model galaxy, lensed with a
magnification of $\mu\geq10$ and of $2<\mu<5$, with [C{\sc
  ii}]/$L_{\rm FIR}$ plotted against CO$(J=1-0)/L_{\rm FIR}$. This
diagram has been used to estimate the density and ionizing radiation
environment in star-forming galaxies (e.g. Stacey et al. 2010,
Valtchanov et al. 2011).  The $500\,{\rm \mu}$m luminosity is taken as a
proxy for the far-infrared (FIR) luminosity $L_{\rm FIR}$ in the
differentially-magnified simulated galaxies. It is clear that
differential magnification has surprisingly little effect on this
diagnostic plot for 
{\bfreferee 
lensed galaxies. For nearly all lens configurations, differential
magnification confines the source to a line almost parallel to the
constant density contours. However, the vertical positioning of this
track was found to depend on the random positioning of the GMCs in the
source model. 
}

If one chooses to use the CO$(J=4-3)$ line as a proxy for the
CO$(J=1-0)$ line, the smaller emitting regions of CO$(J=4-3)$ 
{\bfreferee 
slightly}
changes
the differential magnification effects, as shown also in
Fig.~\ref{fig:cii}. Table \ref{tab:cii} gives the dispersions in the
logarithmic bolometric fractions of the emission lines in
Fig.~\ref{fig:cii}. 
The
dispersion in $\log_{10}{\rm CO}/L_{\rm FIR}$ is smaller than the
measurement errors of {\it H}-ATLAS galaxy ID81, while the standard
deviation of $\log_{10}[$C{\sc ii}$]/L_{\rm FIR}$ is comparable to the
measurement errors for ID81 (Valtchanov et al. 2011).

\begin{table}
\begin{tabular}{lll}
Ratio & $2<\mu<5$ & $\mu\ge 10$ \\
\hline
$\log_{10}($CO$(J=1-0)$/FIR$)$ & 0.16 & 0.10\\
$\log_{10}($CO$(J=4-3)$/FIR$)$ & 0.15 & 0.10\\
$\log_{10}($[C{\sc ii}]/FIR$)$ & 0.17 & 0.19\\
\end{tabular}
\caption{\label{tab:cii} Dispersions in the logarithmic bolometric
  fractions of key emission line diagnostics from Fig.~\ref{fig:cii},
  for the stated magnifications $\mu$.}
\end{table}

The SHINING survey (Survey with {\it Herschel} of the ISM in Nearby
Infrared Galaxies, e.g. Sturm et al. 2010) has
found that the [O{\sc i}] $63\,{\rm \mu}$m emission region in M82 is broadly
co-spatial with that of the [C{\sc ii}] $158\,{\rm \mu}$m emission on
$100-200\,$pc scales (Contursi et al. 2010, Sturm et al. in preparation). If
this is also true of high-redshift star-forming galaxies, the
density-sensitive [O{\sc i}]/[C{\sc ii}] line ratio should be
insensitive to differential magnification.

\subsection{Emission line diagnostics: CO}
Cox et al. (2011) find the $z=4.24$ lensed SMG ID141 has a 
CO Spectral Line Energy Distributions (SLED) similar in shape to that
of the $z=4.05$ SMG GN20 (Carilli et al. 2010) and in the $z=2.5$ lensed
galaxy SMM\,J16359+6612 (Wei\ss\, et al. 2005), though cooler than that of
the $z=5.3$ millimetre-galaxy AZTEC-3 (Riechers et al. 2010). Could
differential magnification skew the CO SLEDs in the lensed galaxies?

Fig.~\ref{fig:cii} already 
{\bfreferee suggests}
the CO
ladder is 
distorted due to the differing spatial extents of the CO emission line
regions in our source model. In this case, the effect on the physical
interpretation is much stronger, because of a much smaller dynamic
range used for the diagnosis of physical conditions. 

\begin{figure}
\centering
\ForceWidth{3.5in}
%\hSlide{-1cm}
%\vspace*{-1cm}
%\BoxedEPSF{h2o.ps}
\BoxedEPSF{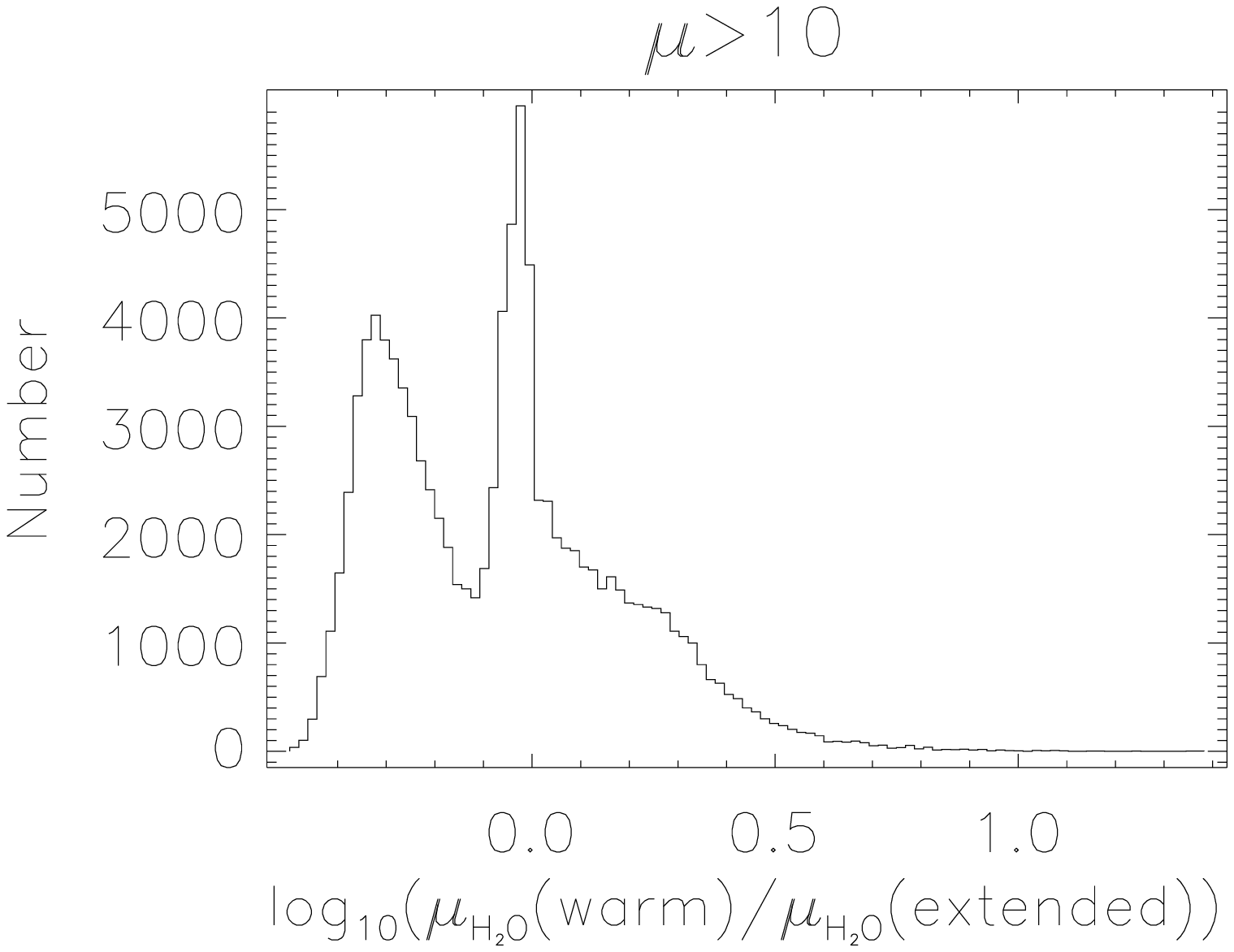}
%\vspace*{-1cm}
\caption{\label{fig:h2o}
Simulated differential magnification of the H$_2$O emission regions
in the $z=2$ SMG model discussed in the text, for a 
$500\,{\rm \mu}$m-selected gravitational lens with a magnification
$\mu>10$. 
}
\end{figure}
\begin{figure*}
\centering
%\ForceWidth{4in}
%\hSlide{-1cm}
%\vspace*{-1cm}
%\BoxedEPSF{ternary_0.ps}
\ForceWidth{3.2in}
\BoxedEPSF{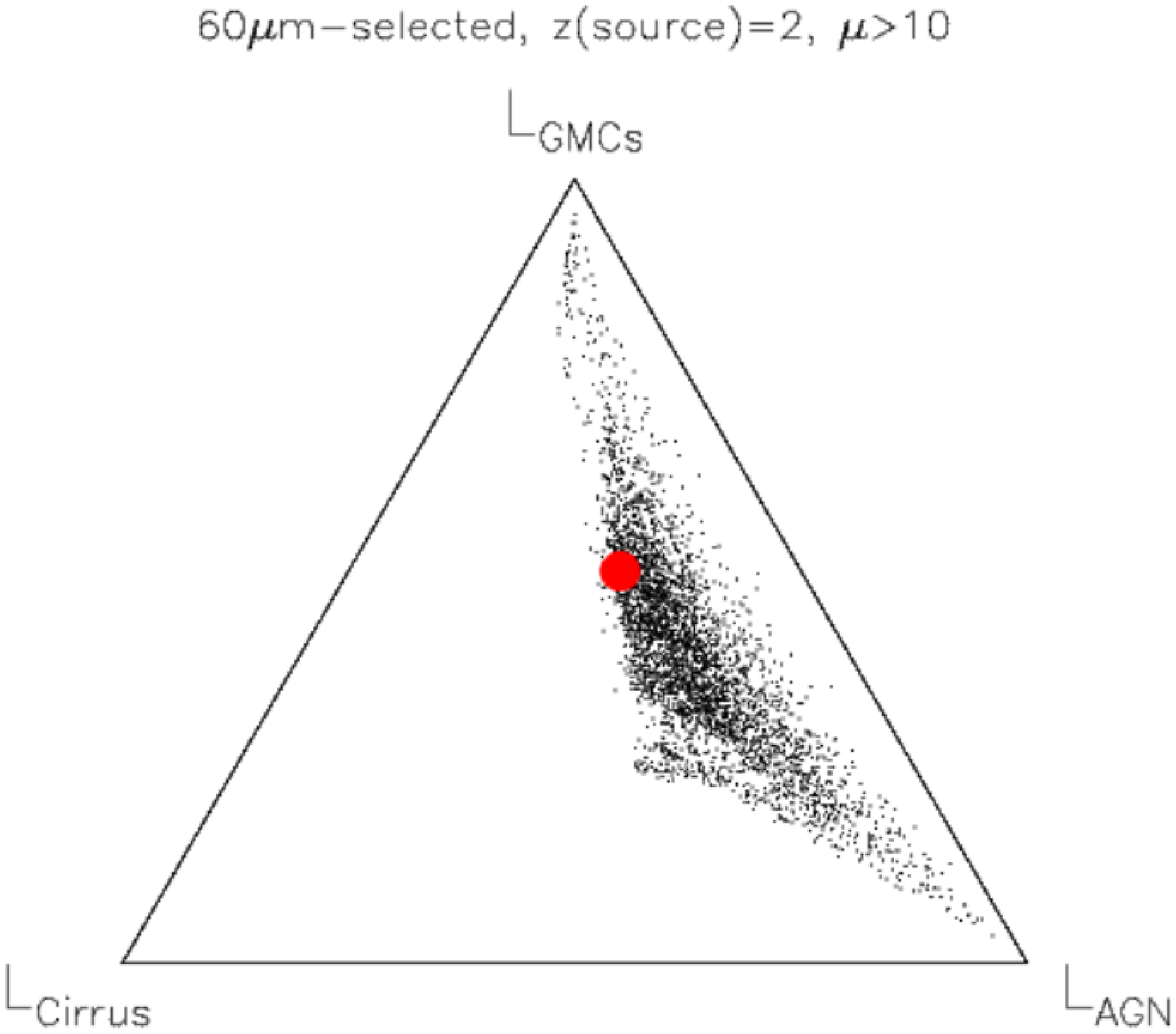}
\ForceWidth{3.2in}
\BoxedEPSF{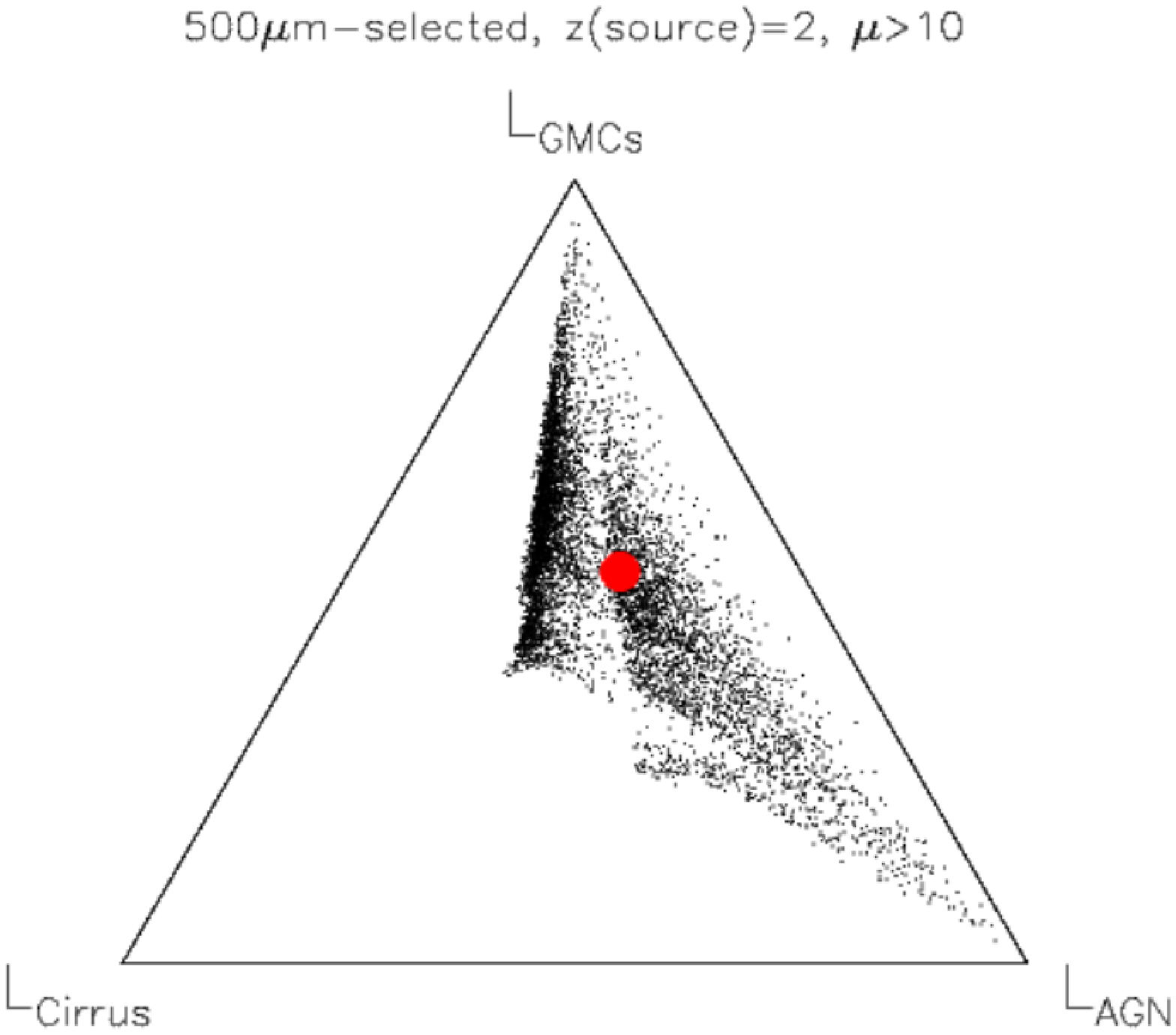}
\ForceWidth{3.2in}
\BoxedEPSF{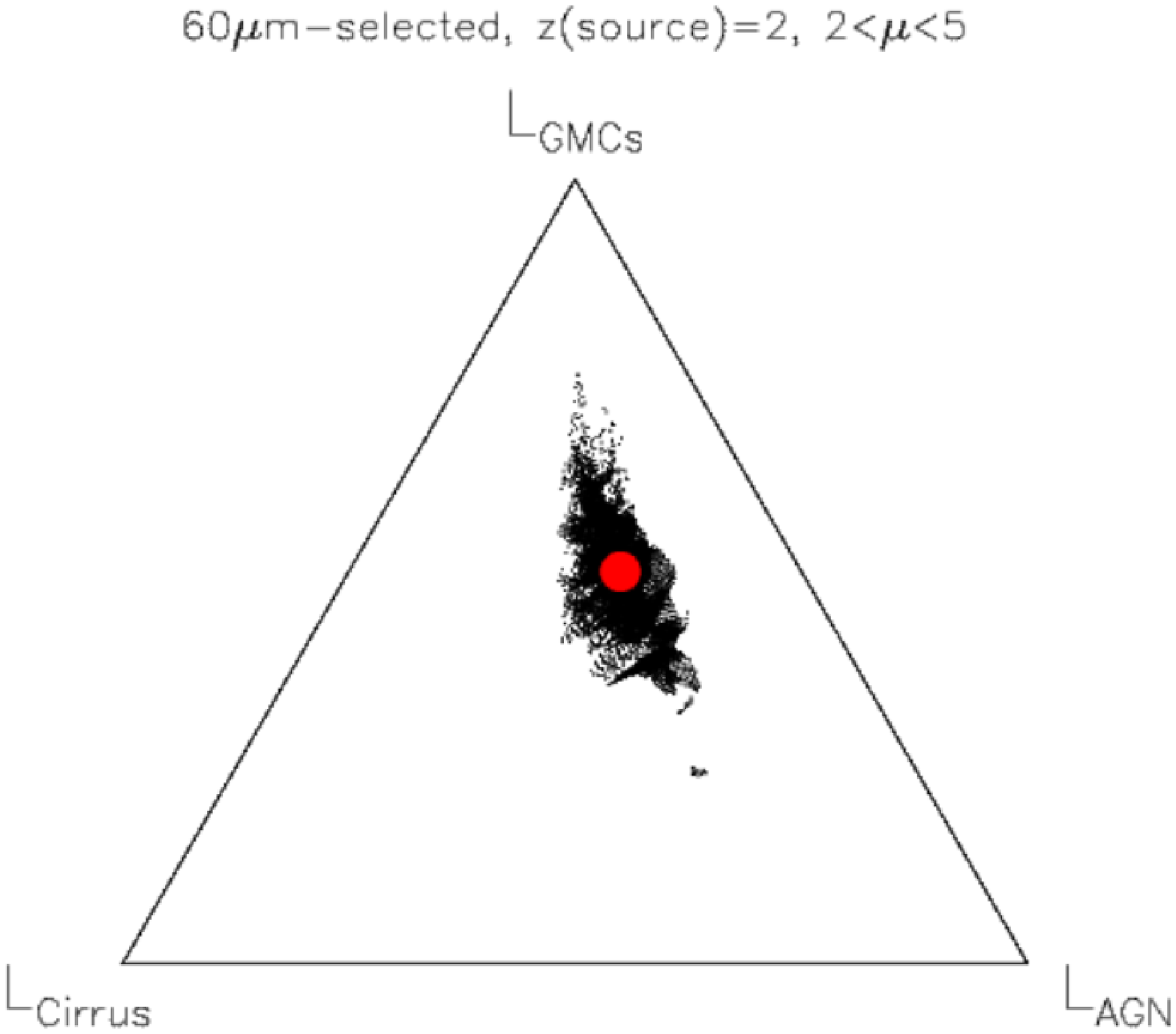}
\ForceWidth{3.2in}
\BoxedEPSF{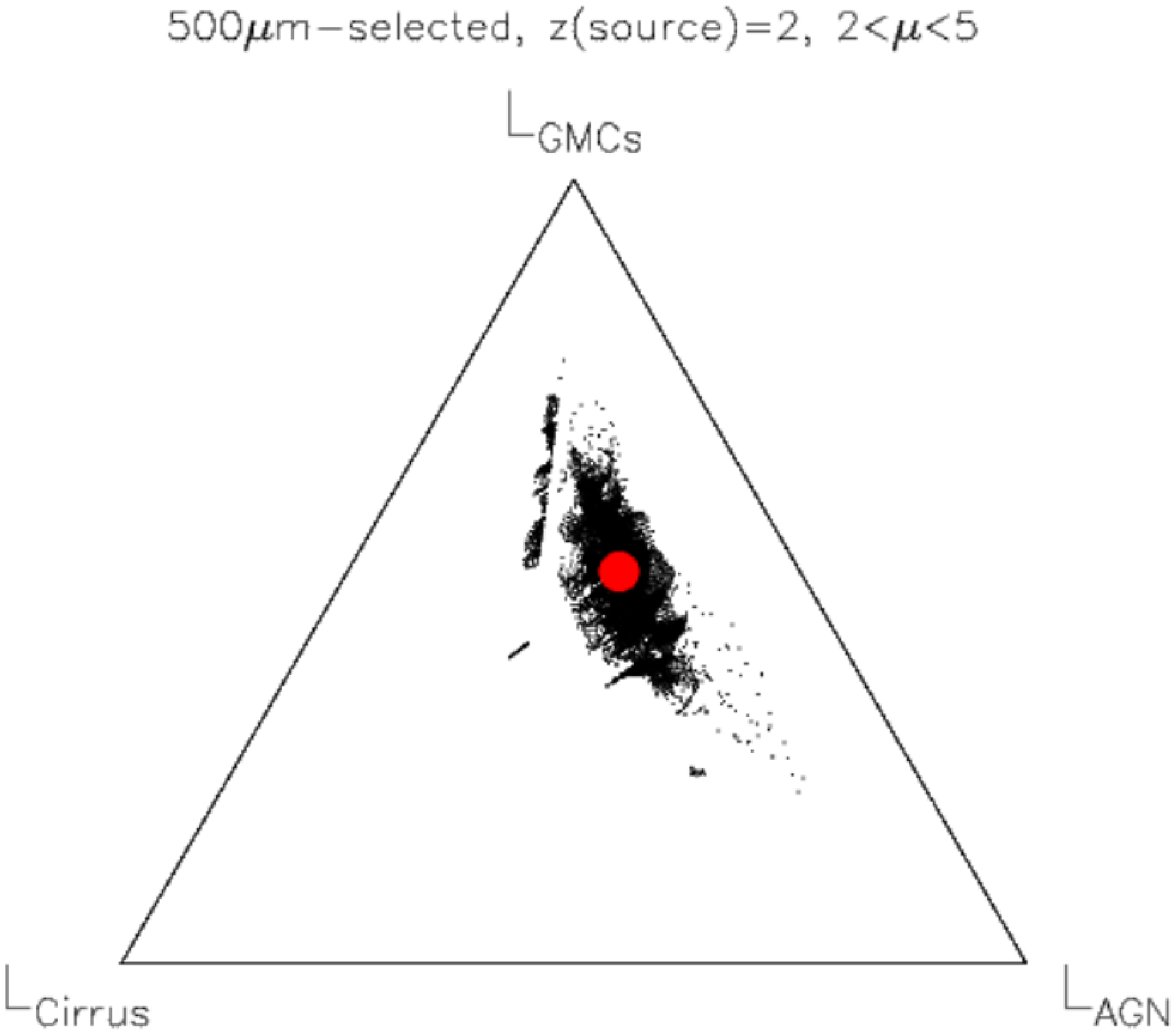}
%\vspace*{-1cm}
\caption{\label{fig:ternary}
Apparent bolometric fractions for the simulated $z=2$ SMG model 
source in Section
\ref{sec:source_model}, gravitationally lensed by the lens population
described in Section \ref{sec:lens_model}, then selected at
$60\,{\rm \mu}$m (left) or $500\,{\rm \mu}$m (right). The upper panels show the
results for high magnification lenses, $\mu>10$; the lower panels show
moderate magnifications lenses, $2<\mu<5$. 
The large red dot represents
the underlying unlensed source, while the small dots represent the
differentially magnified source. The data are represented as a ternary
diagram, i.e. the perpendicular distance from any
side to the opposite corner is proportional to the bolometric
fraction, with $0\%$ on the side and $100\%$ at the corner. Each
corner is labelled with the $100\%$ bolometric fraction at that
point. 
Only a random sparse sample of 2 (10) percent of the configurations are
plotted in the lower (upper) panels, for clarity. 
}
\end{figure*}

Fig. \ref{fig:co_sled} 
shows the SLEDs for the lensed SMG model source. The SLEDs have been
normalised to the CO$(J=1-0)$ point to remove the overall
magnification effects and isolate the differential magnification
effects. Clearly the relative proportions of the cool and warm
molecular gas components vary strongly, according to the random
differential magnification, even for moderate magnification
lenses. 
{\bfreferee Both the $500\,{\rm \mu}$m-selected lenses and $60\,{\rm \mu}$m-selected
  lenses suffer strong differential magnification effects, though the
  effects are not identical. There is also a higher-probability SLED
  (the darker bands in Fig. \ref{fig:co_sled}) within the overall
  dispersion that was found to
  depend on the specific random configuration of the GMCs.}
The warm molecular gas component is in
close physical proximity to the star-forming knots in this model, but
this does not necessarily guarantee that the emission from the warm
gas component is boosted relative to that from the cool component,
even when selecting at an observed wavelength of $500\,{\rm \mu}$m. The
$1\sigma$ dispersion in the apparent CO$(J=6-5)$ line (once normalised
to CO$(J=1-0)$ as in Fig.~\ref{fig:co_sled}) is $32\%$ of the mean
value for $60\,{\rm \mu}$m-selected strong lenses ($\mu>10$), or $20\%$ of
the mean value for $500\,{\rm \mu}$m-selected strong lenses. Even at more
moderate magnifications of $2<\mu<5$, the corresponding dispersions
are $29\%$ at $60\,{\rm \mu}$m and $30\%$ at $500\,{\rm \mu}$m.

\subsection{Emission line diagnostics: HNC, HCN, H$_2$O}\label{sec:hcn}
Omont et al. (2011) found the bolometric fraction of an H$_2$O line in 
the $z=2.3$ lensed SMG SDP\,17b to be higher than that in Mrk\,231,
suggestive of a luminous AGN. Van der Werf et al. (2011) detected four
rotational H$_2$O transitions in the $z=3.9$ lensed quasar
APM\,08279+5255, deriving a gas temperature of $105\pm21$\,K. 
Lupu et al. (2011) reported early results from high density gas
tracers in lensed SMGs, such as HCN, HNC HCO$^+$ and $^{13}$CO. 
Are these observations sensitive to the effects of differential
magnification? 

The $J=1-0$ HNC and HCN transitions at rest-frame wavelengths of
around $3\,$mm have very similar critical densities ($3\times
10^6$\,cm$^{-3}$) and so there is a reasonable expectation for these
lines to be co-spatial.  The HCN:HNC flux ratio is a good diagnostic
of whether the emission line gas is from a photodissociation region
(PDR) or an X-ray dissociation region (XDR), e.g. Loenen et
al. 2007. The HCN:HCO$^+$ and HNC:HCO$^+$ ratios, conversely, are
sensitive to densities in PDRs, since HCO$^+$ has a slightly lower
critical density of $3\times10^5$\,cm$^{-3}$.  Aalto et al. (2009)
find that the $J=3-2$ HCN and HNC emission in Arp\,220 are both
confined to a region $<350$\,pc wide, with most within the central
$30$\,pc. The HCN and HNC emission has therefore been pessimistically
modelled in the SMG model source as point sources separated by
$80$\,pc. The simulations confirmed that the differential
magnification effect are essentially negligible between these two
emission regions.

A similar optimistic situation exists for {\it moderate} magnification
in H$_2$O lines.  Following the precedent of Mrk\,231, the H$_2$O
emission is modelled as two components, one with a radius of
$120\,$pc, and the other with a major axis of $1$\,kpc (Section
\ref{sec:source_model}). These components contribute comparable flux
in Mrk\,231 at upper energy levels below $200\,$K, while the warm
component dominates at higher temperatures. For $500\,{\rm \mu}$m-selected
lenses with {\it moderate} magnifications of $2<\mu<5$, the
simulations show almost no differential magnification between the two
components. Very similar results were obtained for other source
redshifts.

However, the situation is not as optimistic for H$_2$O lines at
magnifications $\mu>10$. These configurations have strong differential
magnification effects between these two components, shown in
Fig.\,\ref{fig:h2o}. This figure also shows a tail to high
differential magnifications. A similar tail is also present in the
magnification ratio of warm $H_2$O to $500\,{\rm \mu}$m emission, and in
that of the warm $H_2$O to GMC emission. The tail is an alternative
interpretation of the H$_2$O$(2_{02}-1_{11})$ detection in the lensed
{\it Herschel}-ATLAS galaxy SDP.17b (Omont et al. 2011), since in
Mrk\,231 about two-third of the luminosity in this line comes from
this compact warm component.

The ratios of H$_2$O to high-$J$ CO transitions have also been argued
as a useful diagnostic of PDR versus XDR conditions
(e.g. Gonz\'alez-Alfonso et al. 2010, van der Werf et al. 2010). In
highly magnified lensed systems this will clearly be contingent on the
degree that the emissions can be treated as co-spatial.

\begin{figure}
\centering
\ForceWidth{3.75in}
\hSlide{-1cm}
%\vspace*{-1cm}
%\BoxedEPSF{agn_lbol_histogram.ps}
\BoxedEPSF{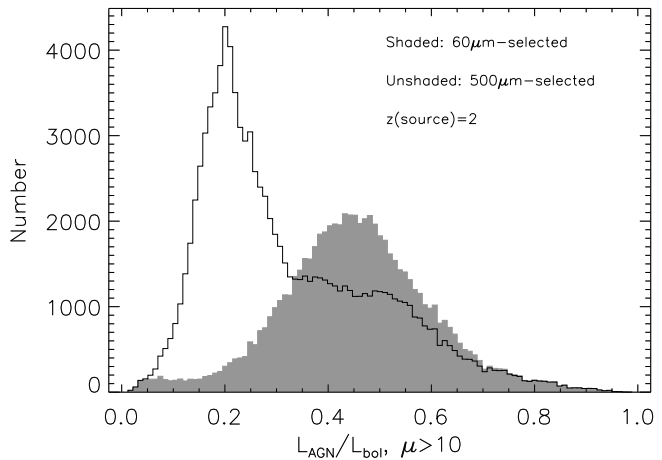}
%\vspace*{-1cm}
\caption{\label{fig:histogram} Histogram of apparent AGN bolometric
  luminosity fractions in the simulated $z=2$ SMG model source
  from Section \ref{sec:source_model}, and a lens population as described
  in Section \ref{sec:lens_model}, for lenses with magnification
  $\mu\geq10$ selected at an observed frame of $60\,{\rm \mu}$m and
  $500\,{\rm \mu}$m. The underlying AGN bolometric fraction is $0.3$ in this
  model (Section \ref{sec:source_model}). On average, the $60\,{\rm \mu}$m selection
  over-represents configurations in which the AGN is close to a
  caustic and has a consequent strong differential
  magnification. Conversely, $500\,{\rm \mu}$m selection more often has GMCs
  with the greater magnification. Nevertheless, note that many lens
  configurations that satisfy the $60\,{\rm \mu}$m lens selection would also
  satisfy the $500\,{\rm \mu}$m lens selection in this simulation.}
\end{figure}
\begin{figure}
\centering
\ForceWidth{3.75in}
\hSlide{-1cm}
%\vspace*{-1cm}
%\BoxedEPSF{main_sequence.ps}
\BoxedEPSF{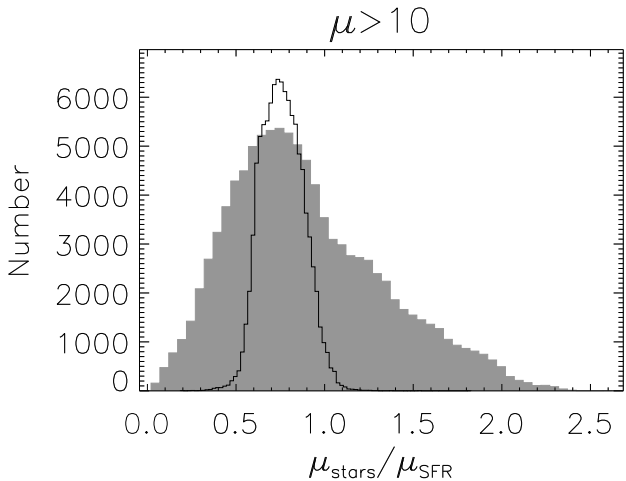}
%\vspace*{-1cm}
\caption{\label{fig:main_sequence} 
The shaded histogram (bin size $0.05$) 
shows the relative magnification factor of the
old stars component with that of
the star forming regions in the $z=2$ SMG model source with
magnifications $\mu>10$. The 
unshaded histogram (bin size $0.02$) 
gives the result when using the $500\,{\rm \mu}$m luminosity as a proxy for
the star formation rate. Both histograms are for a simulated
$500\,{\rm \mu}$m-selected lens. 
}
\end{figure}

\subsection{Apparent bolometric
  fractions}\label{sec:bolometric_fractions}
Broad-band SEDs have long been used to derive the relative bolometric
fractions in galaxies of AGN and star formation, via radiative
transver models (e.g. Rowan-Robinson 1980, 2000; 
Rowan-Robinson \& Crawford 1989; 
Granato, Danese \& Franceschini 1996; 
Green \& Rowan-Robinson 1996; 
Silva et al. 1998, 2011; 
Efstathiou, Rowan-Robinson \& Siebenmorgen 2000, 2006; 
Popescu et al. 2000; 
Misiriotis et al. 2001;
Farrah et al. 2002; 
Verma et al. 2002; 
Tuffs et al. 2004; 
M\"{o}llenhoff, Popescu \& Tuffs 2006; 
Popescu et al. 2011). However, it
is rare that differential magnification is explicitly incorporated
into SED fits of lensed star-forming galaxies (e.g. Deane et
al. 2011).

It is very clear from the lensed SMG model that 
differential magnification has a strong effect on the apparent
bolometric fractions of the continuum components in the model
source. Fig.~\ref{fig:ternary} shows the
effect of differential magnification on these components in the SMG
model source, expressed as
a ternary diagram. At each apex, the bolometric fraction of that
component is $100\%$. Along each side, the bolometric fraction of the
opposite component is $0\%$. For example, in the bottom-left corner,
the bolometric fraction of the cirrus component would be $100\%$,
while all along the right-hand side the cirrus bolometric fraction is
$0\%$. The bolometric fraction is proportional to the perpendicular
distance from a side.

\begin{figure*}
\ForceWidth{3in}
\BoxedEPSF{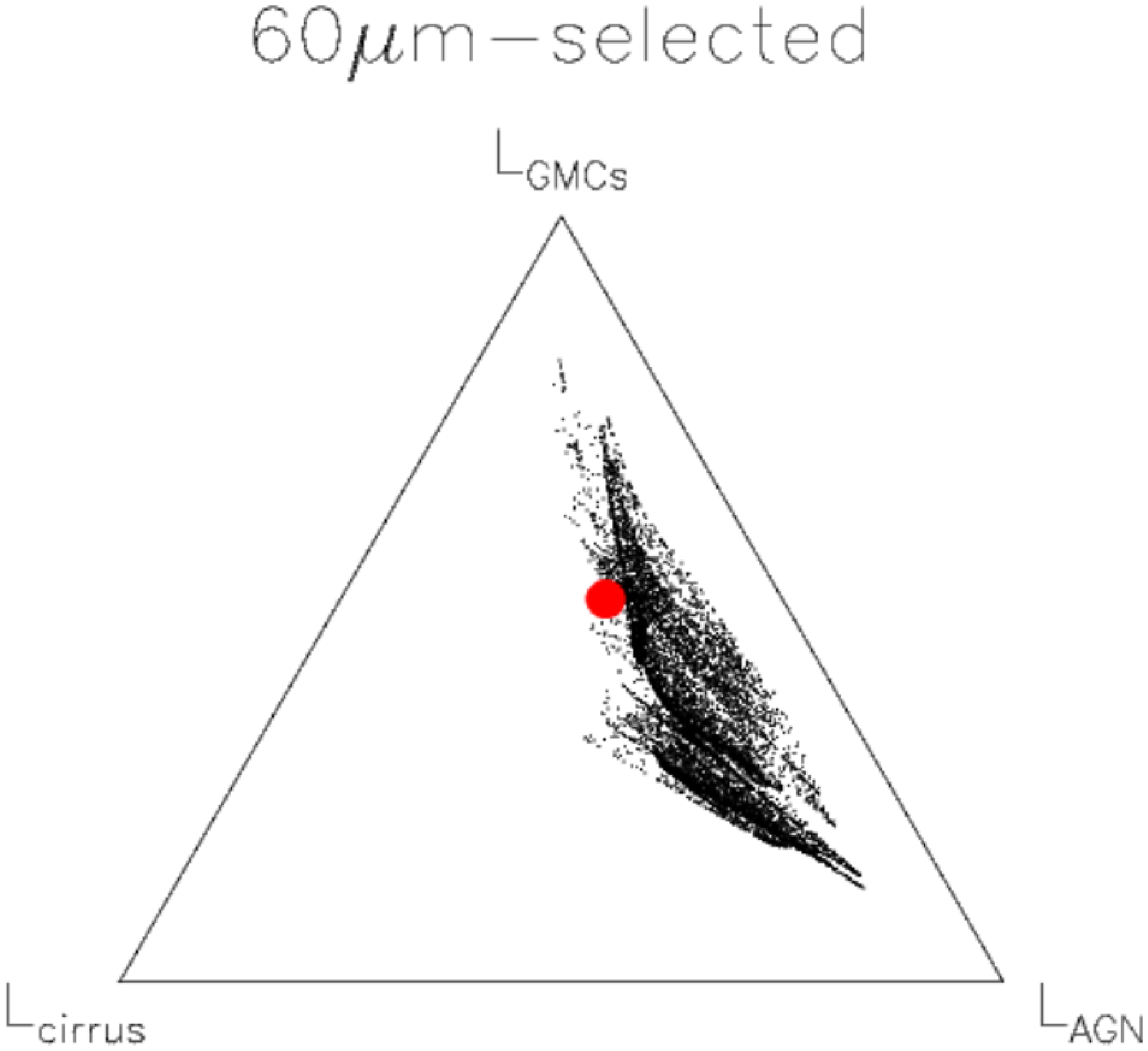}
\ForceWidth{3in}
\BoxedEPSF{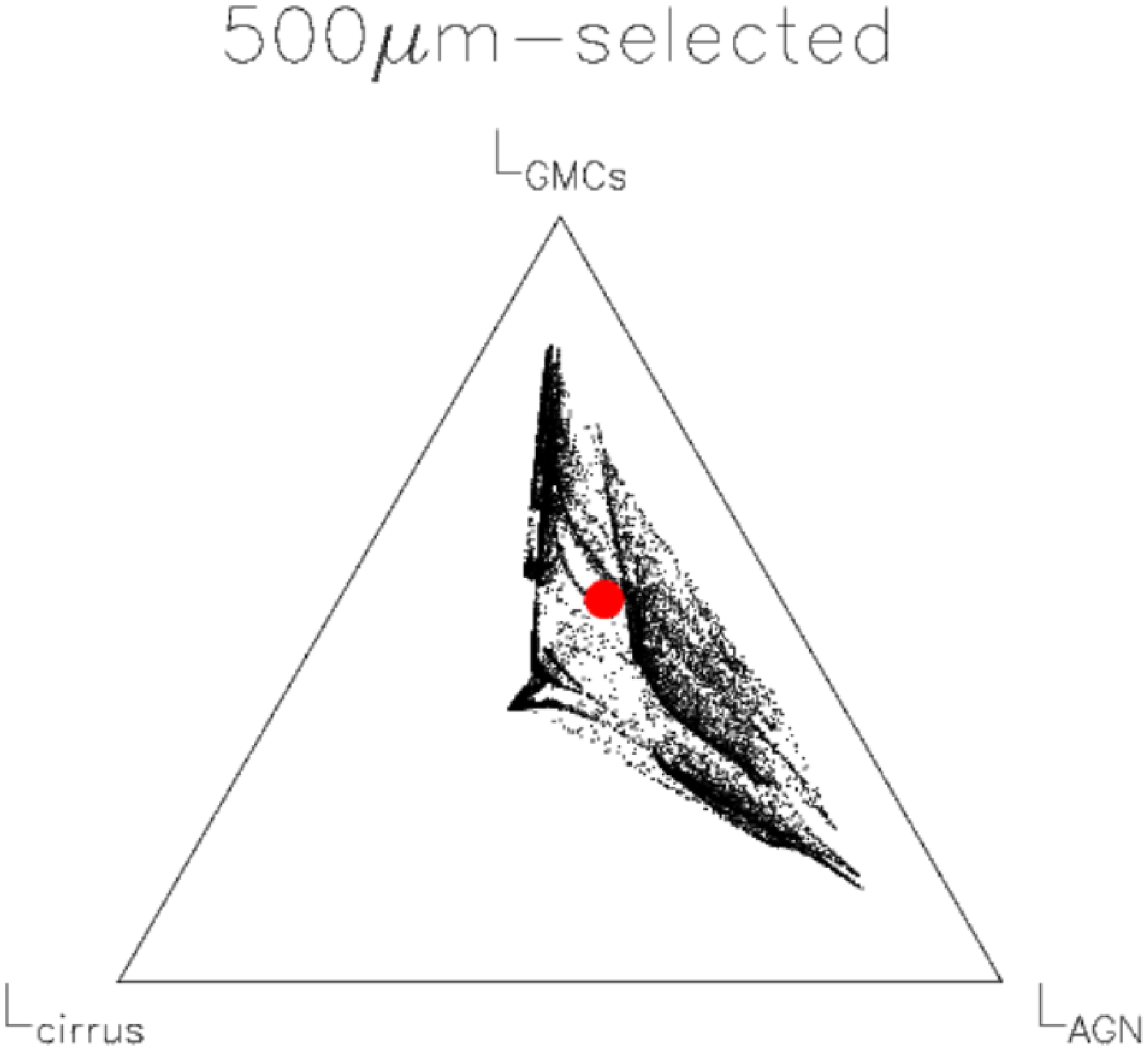}
\ForceWidth{3in}
\BoxedEPSF{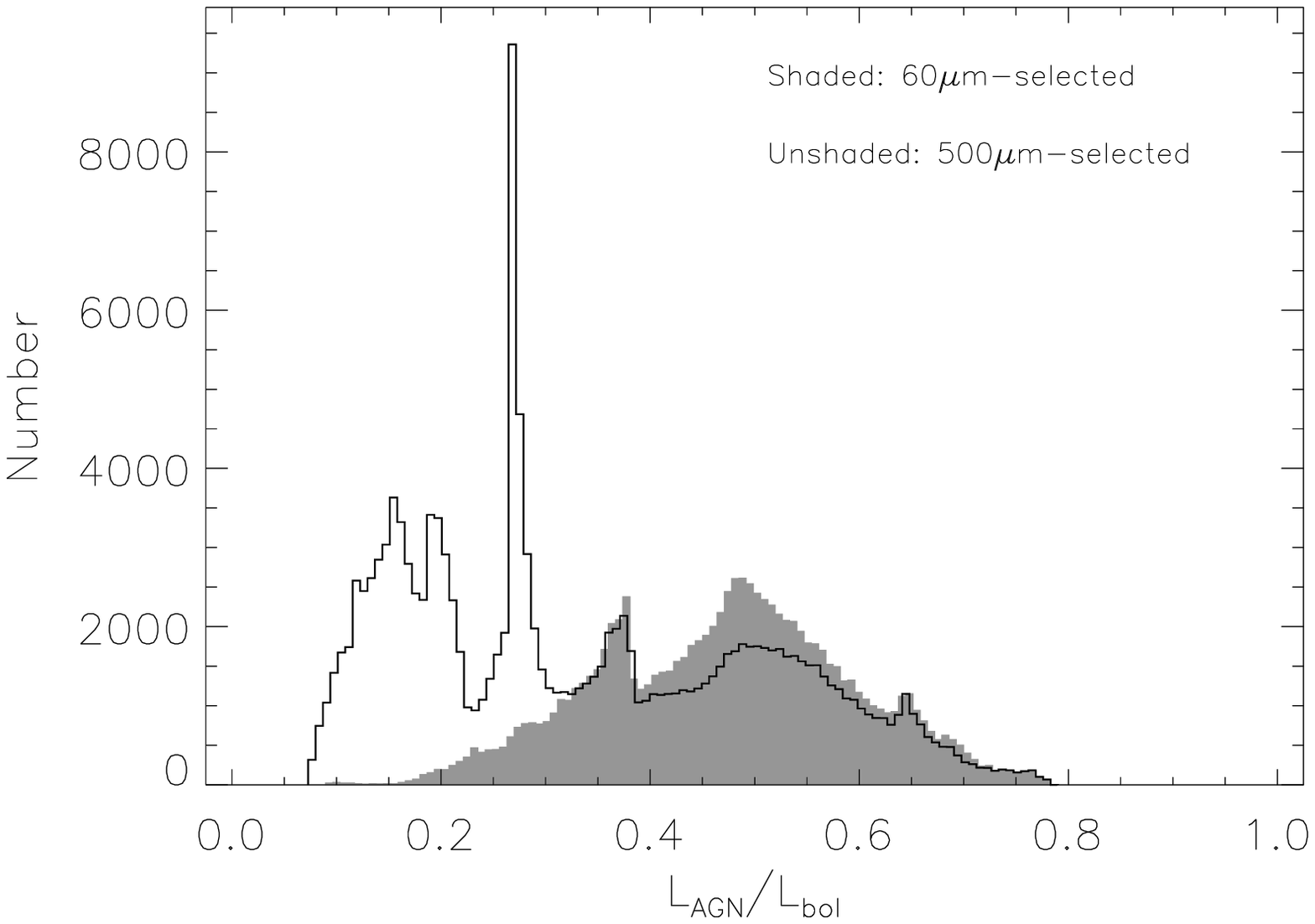}
\ForceWidth{3in}
\BoxedEPSF{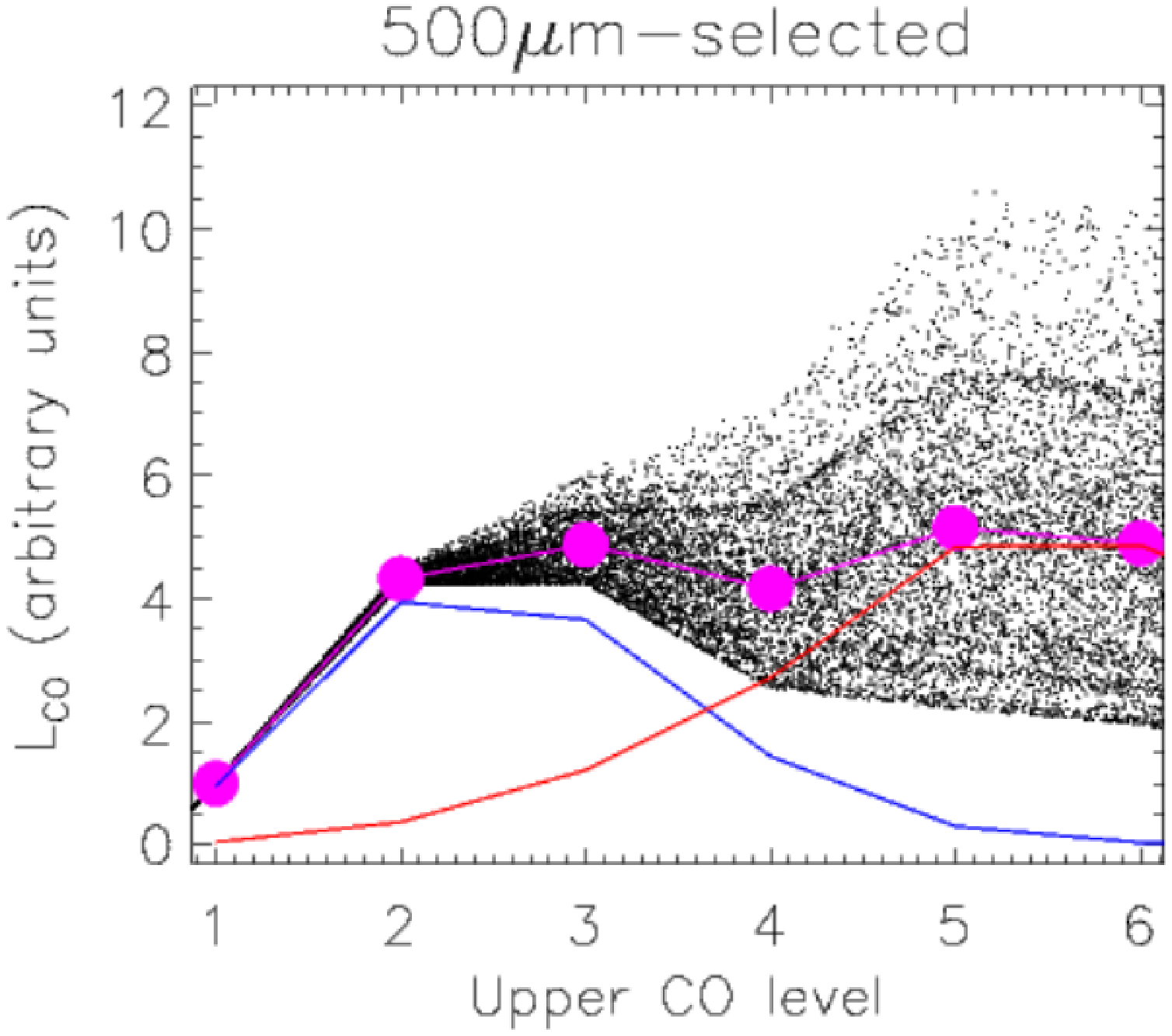}
\caption{\label{fig:f10214} A selection of results for simulated high
  magnification ($\mu>10$) lensing configurations for the source and
  lens redshifts of IRAS\,FSC\,10214+4724. Symbols as in
  Figs. \ref{fig:co_sled}, \ref{fig:ternary} and \ref{fig:histogram}.}
\end{figure*}

The large red dots in Fig.~\ref{fig:ternary} 
represents the underlying bolometric fractions in the unlensed source,
from Section \ref{sec:source_model}. The small dots represent the
simulated magnified source, for $60\,{\rm \mu}$m-selected 
lenses and for $500\,{\rm \mu}$m-selected
lenses. The cirrus
does not have a strong differential magnification, because its
physical size is large compared to the caustic sizes in the source
plane. However, the GMCs and the AGN are small, and therefore subject
to strong differential magnification effects. 

In the case of high magnifications ($\mu>10$), these differential
magnification effects are very large.  It is clear from
Fig.~\ref{fig:ternary} that without a lens model to correct for
differential magnification in strongly lensed systems, the observed
bolometric fractions of AGN, starburst and cirrus are so unreliable as
to be useless for diagnosing the underlying bolometric fractions.

A curious feature of Fig.~\ref{fig:ternary} are the unoccupied regions
for strong lenses ($\mu>10$). The spatial extent of the cirrus limits
its maximum magnification, which corresponds to a diagonal threshold
parallel to the right-hand side. However, there are clearly other
thresholds also at work: for example, there are no simulated lens
configurations with a maximum cirrus bolometric fraction {\it and} an
AGN bolometric fraction of $\sim0\%$. This reflects the specific
spatial configuration of the background source, i.e. there are no
configurations in which a GMC is close to a caustic while the AGN is
simultaneously distant from all caustic curves.

The different distributions in Figs.~\ref{fig:ternary} 
demonstrate that the selection wavelength imposes
a strong bias on the observed apparent bolometric fractions at high
magnification factors.  These
differences between the observed-frame selection wavelengths are
clearer in Fig.~\ref{fig:histogram}, which shows the AGN bolometric
fraction for the two selection wavelengths for $\mu>10$ lenses. 
Selecting at $60\,{\rm \mu}$m is
sampling $<20\,{\rm \mu}$m in the rest-frame, so one may expect
$60\,{\rm \mu}$m selection to favour configurations the AGN torus is close
to a caustic in the source plane. This is indeed what is seen in
Fig.~\ref{fig:histogram}. This also naturally explains the lens
configuration of IRAS F10214$+$4724, in which a cusp appears to lie
across the AGN dust torus (Lacy, Rawlings and Serjeant 1998). A
similar suggestion was made for other lensed ultraluminous galaxies by
Blain (1999).  Conversely, $500\,{\rm \mu}$m selection favours the
differential magnification of the most luminous small sources at
$500\,{\rm \mu}$m, which are the GMCs. Note however that in this simulation,
many lens--source configurations that would satisfy the
$60\,{\rm \mu}$m-based lens selection, would also satisfy the
$500\,{\rm \mu}$m-based lens selection; it is simply that there are
additional configurations that are are more common in the $500\,{\rm \mu}$m
case (Fig.~\ref{fig:histogram}). 

For moderate 
{\bfreferee magnification}
lenses ($2<\mu<5$), the observed AGN
fraction {\it is} an unbiased estimator of the underlying value. In
these simulations, the apparent AGN bolometric fraction for both
$60\,{\rm \mu}$m-selected and $500\,{\rm \mu}$m-selected galaxies is roughly
normally distributed about the underlying value of $0.3$, with a
dispersion of $0.03$.

\subsection{The galaxy ``main sequence''}
The suggestion has been made that star-forming galaxies have a
characteristic specific star formation rate, which evolves from
$z=0-2$ and then plateaus (e.g. Noeske et al. 2007, Elbaz et al. 2007,
Daddi et al. 2007, Stark et al. 2009, Gonz\'alez et al. 2010). The
scatter in this relationship is around a factor of two or more. There
has been some discussion as to whether submillimetre-selected galaxies are
typical of the population as a whole, or whether they are outliers in
this relation (e.g. Tacconi et al. 2006, 
{\bfreferee Hainline et al. 2011,} 
Micha{\l}owski et
al. 2012). Could differential magnification drive a lensed submillimetre
galaxy into appearing as an outlier? 

This effect would be strongest at high magnifications, so the $\mu>10$
case will be discussed first. 
Selecting the SMG model lenses at $500\,{\rm \mu}$m, and assuming that the
old stellar population can be observed at a wavelength at which AGN
contributions are negligible and that the foreground lens can be
subtracted (e.g. Hopwood et al. 2011), the observed flux ratio of the old
stellar population to that of the star-forming regions varies by
$\pm0.26$\,dex ($1\sigma$), as shown in
Fig.\,\ref{fig:main_sequence}. 
(This also assumes that one can successfully address problems of the 
choice of initial mass function, evolutionary synthesis models, and
star formation histories in estimating the stellar masses; 
e.g. Micha{\l}owski et al. 2012.) 
The median differential magnification
ratio is $0.86$. Using the $500\,{\rm \mu}$m emission itself as a proxy for
the star formation rate, the dispersion reduces to $\pm0.07$ with a
median differential magnification ratio of $0.77$
(Fig.\,\ref{fig:main_sequence}).  

For more modest magnifications ($2<\mu<5$), the effects are much
smaller. The distributions are centred on $\mu_{\rm stars}/\mu_{\rm
  SFR}=1$ with a dispersion of $0.14$, or $0.06$ if using $500\,{\rm \mu}$m
luminosity as a proxy for the star formation rate. 

There are two lessons to be drawn from this. Firstly, the specific
star formation rates can be misleading if the star formation rate
proxy is also the flux at the selection wavelength.  This is closely
related to the fact that bolometric fractions from the SED shape can
be misleading, as noted in the previous section. Secondly, the
dispersion induced by differential magnification in this case is
comparable to the rather broad dispersion in the galaxy ``main
sequence''. A large offset from this correlation 
%(as in e.g. Tacconi
%et al. 2006) 
would therefore be unlikely to be attributable to
differential magnification. 

\subsection{Results for an IRAS\,FSC\,10214+4724-type lens}\label{sec:f10214}
{\bfreferee How sensitive are the simulations to the angular resolution?}
The simulation was repeated 
{\bfreferee a resolution of $0.001$ arcsec, i.e. ten
times finer than the lens population above,}
for a lens at a fixed source redshift of
$z=2.286$ and a lens at $z=0.9$, 
modelled on the example of
IRAS\,FSC\,10214+4724. The orientation of the lens ellipticity was
fixed at an arbitrary value of ${\rm \pi}/4$ relative to that of the
source. The maximum magnification
at an observed wavelength of $500\,{\rm \mu}$m was $17$, while at an
observed wavelength of $60\,{\rm \mu}$m it was $52$.  Fig.~\ref{fig:f10214}
shows a selection of results for this case.  The results are
qualitatively similar to the $z=2$ lens case with a constant comoving
population of lenses; the peaks in the AGN bolometric fraction
histograms are due to particular alignments of source structure with
cusps in this model.  No significant quantitative differences in
Fig.~\ref{fig:f10214} were found with an otherwise identical
simulation at $0.01$ arcsec resolution, justifying the resolution choice of
the lens population simulations above, despite the fact that the
smallest source structures had been treated as de facto point sources.
More results from this lens configuration are found in Serjeant
(2011).

{\bfreferee 
\subsection{Results for other wavelengths and source redshifts: ALMA
  and SMA}\label{sec:various_zsource}
How does differential magnification depend on source redshift and
observed wavelength, and how can submillimetre/millimetre-wave
continuum photometry and interferometry illuminate the bolometric
fractions?

The discussion up to this point has been restricted to a source
redshift of $z=2.0$, with the expectation from Fig.\,\ref{fig:thetae}
that there should not be a strong dependence on the source
redshift, with the exception of a possible dependence on the
rest-frame wavelengths being sampled. Confirmation of this can be found in
Fig.\,\ref{fig:various_zsource} in which the apparent AGN bolometric fraction
is plotted for source redshifts of $z=1$ and $z=4$. 

Most of the results in this paper for $500\,{\rm \mu}$m-selected galaxies
apply equally to $1.4$\,mm-selected galaxies, since both wavelengths
usually lie well within the Rayleigh-Jeans tail. However, at
sufficiently high redshifts, this situation must
change. Fig.\,\ref{fig:magnification_wavelength} presents a comparison
of the monochromatic magnifications at $500\,{\rm \mu}$m and $1.4$\,mm for
source redshifts of $z=1$, $z=2$ and $z=4$. Only at the highest
redshifts and high magnifications are there appreciable differences in
the monochromatic magnification factors. A corollary is that {\it
  e.g.} $870\,{\rm \mu}$m Sub-Millimeter Array (SMA) follow-ups of
$500\,{\rm \mu}$m-selected lenses will be accurately recovering the
$500\,{\rm \mu}$m magnifications and morphologies, provided the $500\,{\rm \mu}$m
passband is not close to the peak of the SED, or that the
magnification factor is not large. Similarly, SMA mapping will almost
always recover the $1.4$\,mm morphologies and magnifications. ALMA
mapping of submillimetre/millimetre-wave-selected strongly lensed
galaxies will also recover the morphologies and magnifications at the
selected wavelength, provided that both the ALMA band and the
selection wavelength are on the Rayleigh-Jeans tail (or that the ALMA
band and the selection wavelengths are similar).

However, it does not necessarily follow that single-band photometry is
a good tracer of any particular component of the background
galaxy. Fig.\,\ref{fig:magnification_components} correlates the
monochromatic magnification factors at selected wavelengths with the
magnifications of the AGN, starburst and cirrus components. As one
might expect, mid-infrared rest-frame photometry is a good tracer of
the magnification of the AGN component. The magnification of the
cirrus component is well-traced by photometry on the Rayleigh-Jeans
tail. The starburst component, however, appears not to be well-traced
by any single monochromatic photometric measurement. In order to
decouple the starburst and cirrus components, one must either obtain
more extensive coverage of the FIR to
submillimetre/millimetre-wave SED, or obtain resolved imaging to
decouple the FIR-luminous components.

{\bfreferee 
\begin{figure}
\centering
\ForceWidth{3.5in}
%%\hSlide{-1.8cm}
%%\vspace*{-1cm}
%\BoxedEPSF{cartoon.ps}
\BoxedEPSF{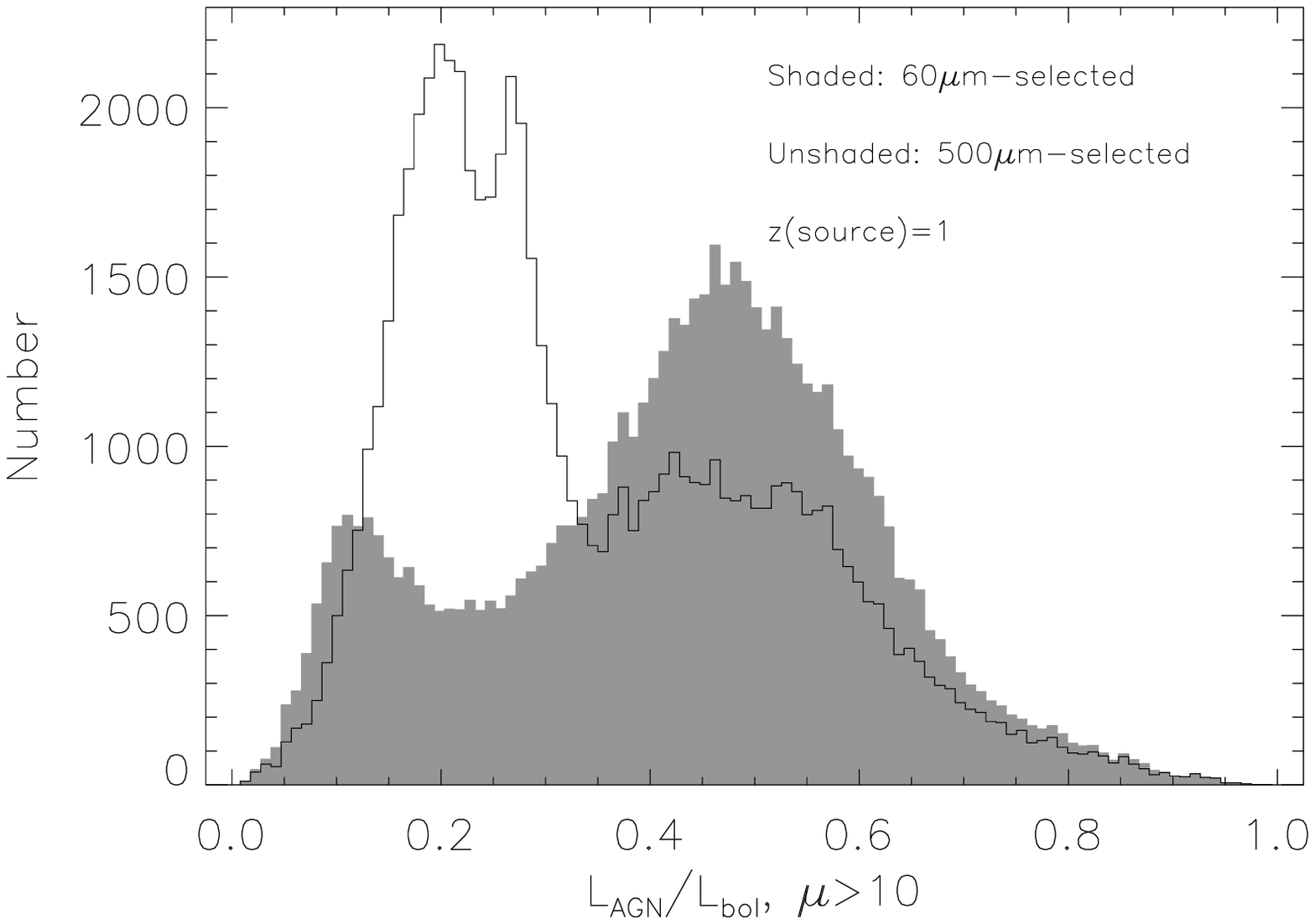}
\ForceWidth{3.5in}
%%\hSlide{-1.8cm}
%%\vspace*{-1cm}
%\BoxedEPSF{cartoon.ps}
\BoxedEPSF{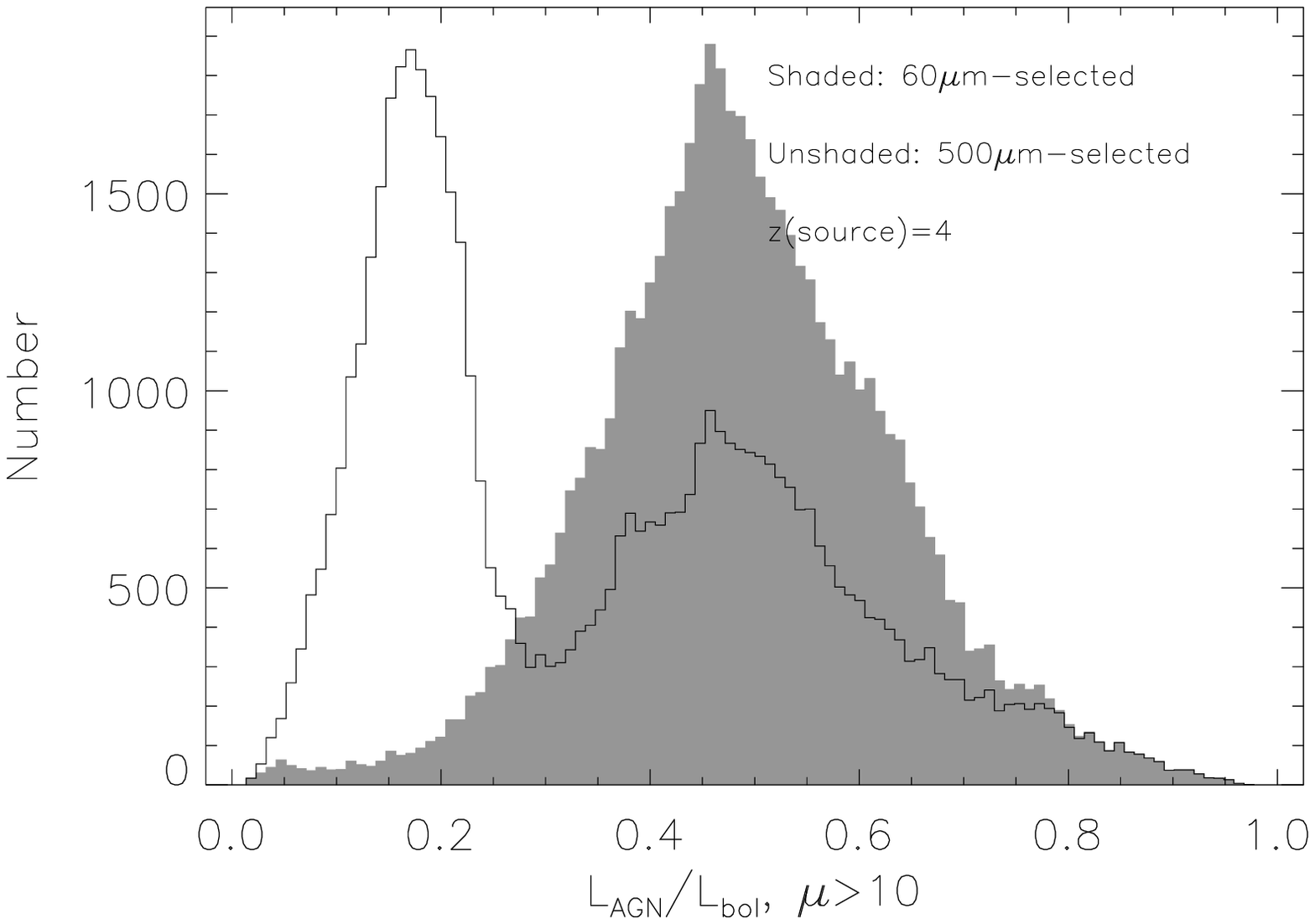}
%%\vspace*{-1cm}
\caption{\label{fig:various_zsource}
The apparent AGN bolometric fractions for a source redshift of $z=1.0$
and $z=4.0$, for $\mu>10$ lensed galaxies selected at $60\,{\rm \mu}$m (grey) and
$500\,{\rm \mu}$m (open). Note the similarity to Fig.\,\ref{fig:histogram}.
}
\end{figure}
}

{\bfreferee 
\begin{figure}
\centering
%\ForceWidth{4.2in}
\ForceWidth{1.32in}
\hSlide{-1.45cm}
\vspace*{-1cm}
\BoxedEPSF{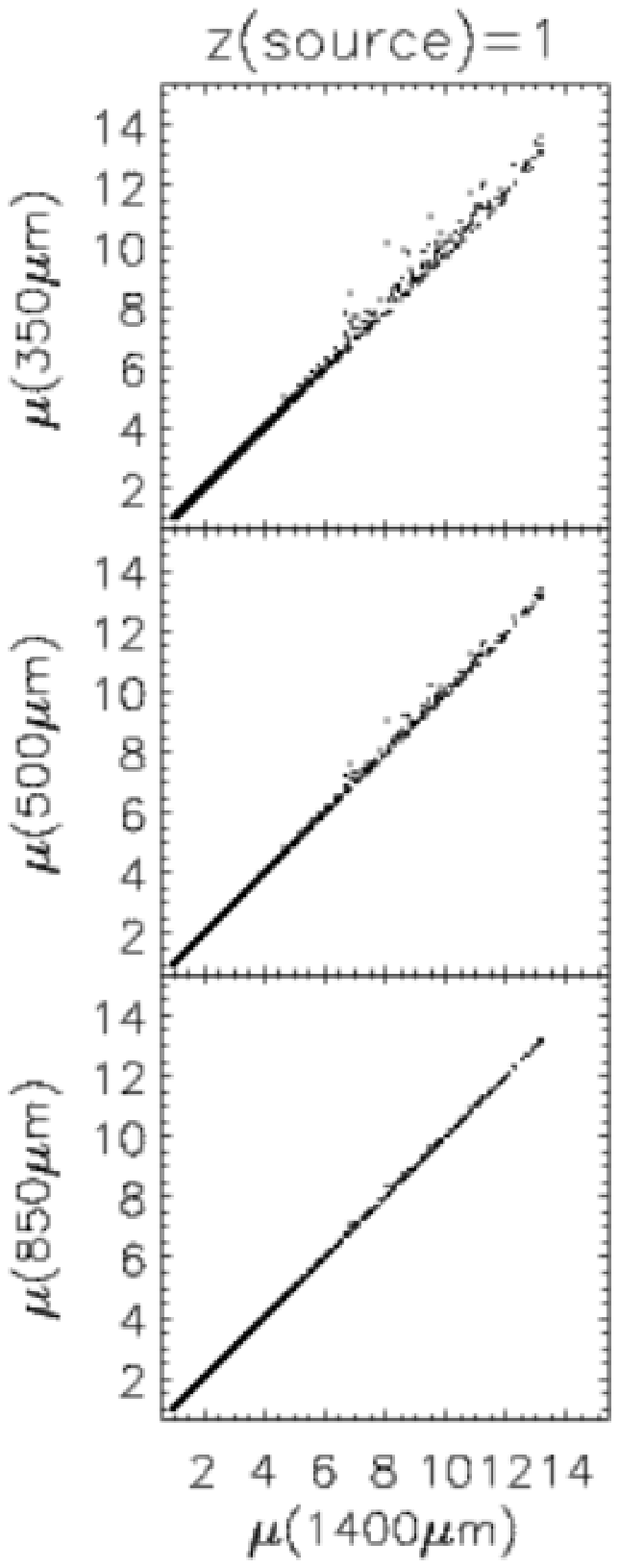}
%\ForceWidth{4.2in}
\ForceWidth{1in}
\hSlide{-1.7cm}
%\vSlide{-7.65cm}
\BoxedEPSF{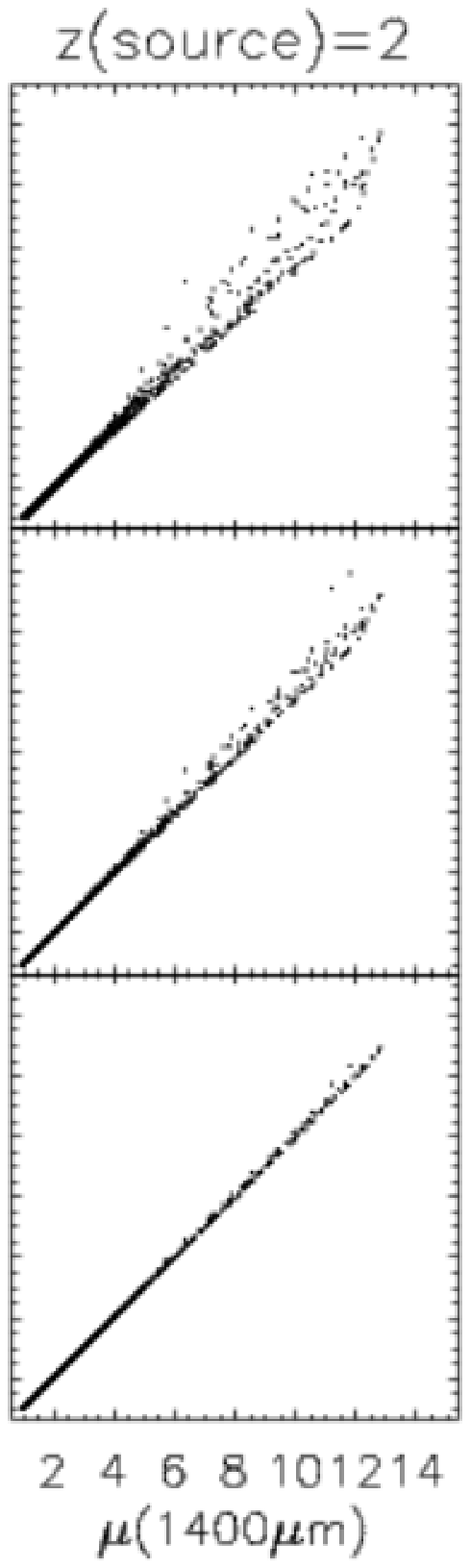}
%\ForceWidth{4.2in}
\ForceWidth{1.025in}
\hSlide{2.8cm}
\vSlide{-8.6cm}
%\vSlide{-15.3cm}
\BoxedEPSF{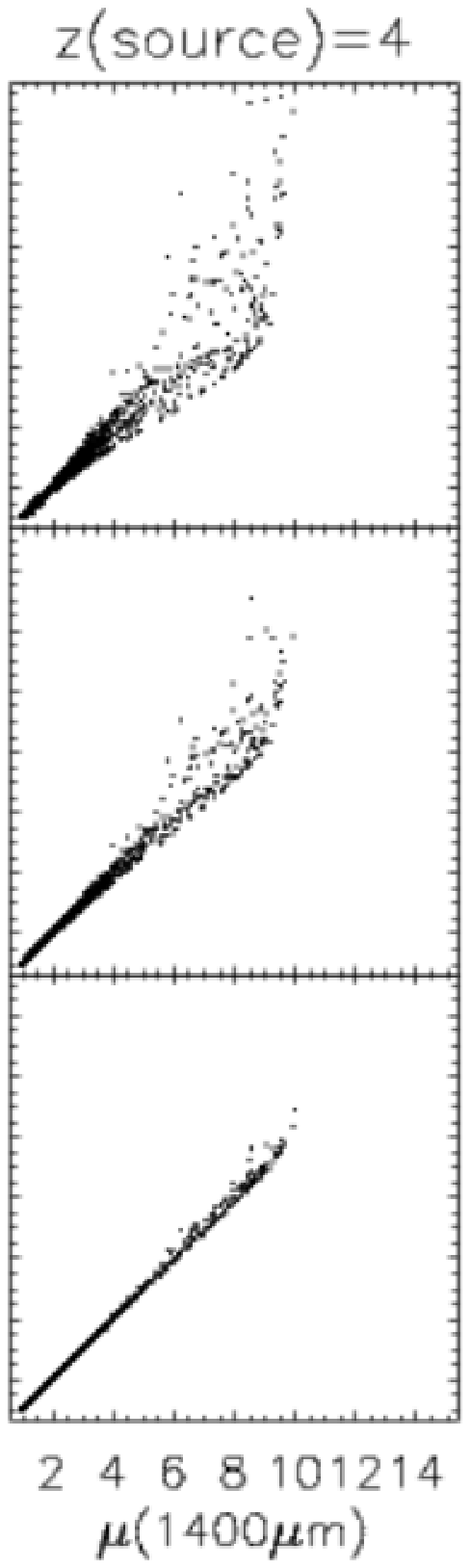}
%\vspace*{-15.3cm}
\vspace*{-8.6cm}
\caption{\label{fig:magnification_wavelength} The monochromatic
  magnification factors at observed wavelengths of $350\,{\rm \mu}$m,
  $500\,{\rm \mu}$m, $850\,{\rm \mu}$m and $1.4$\,mm, for source redshifts of
  $z=1$, $z=2$ and $z=4$ and lens redshift of half the
  source redshift. Note that the magnifications are essentially
  identical until the passband moves off the Rayleigh-Jeans tail.  }
\end{figure}
}

{\bfreferee 
\begin{figure}
\centering
%\ForceWidth{4.2in}
\ForceWidth{1.315in}
\hSlide{-1.45cm}
\vSlide{-0.05cm}
\vspace*{-0.5cm}
\BoxedEPSF{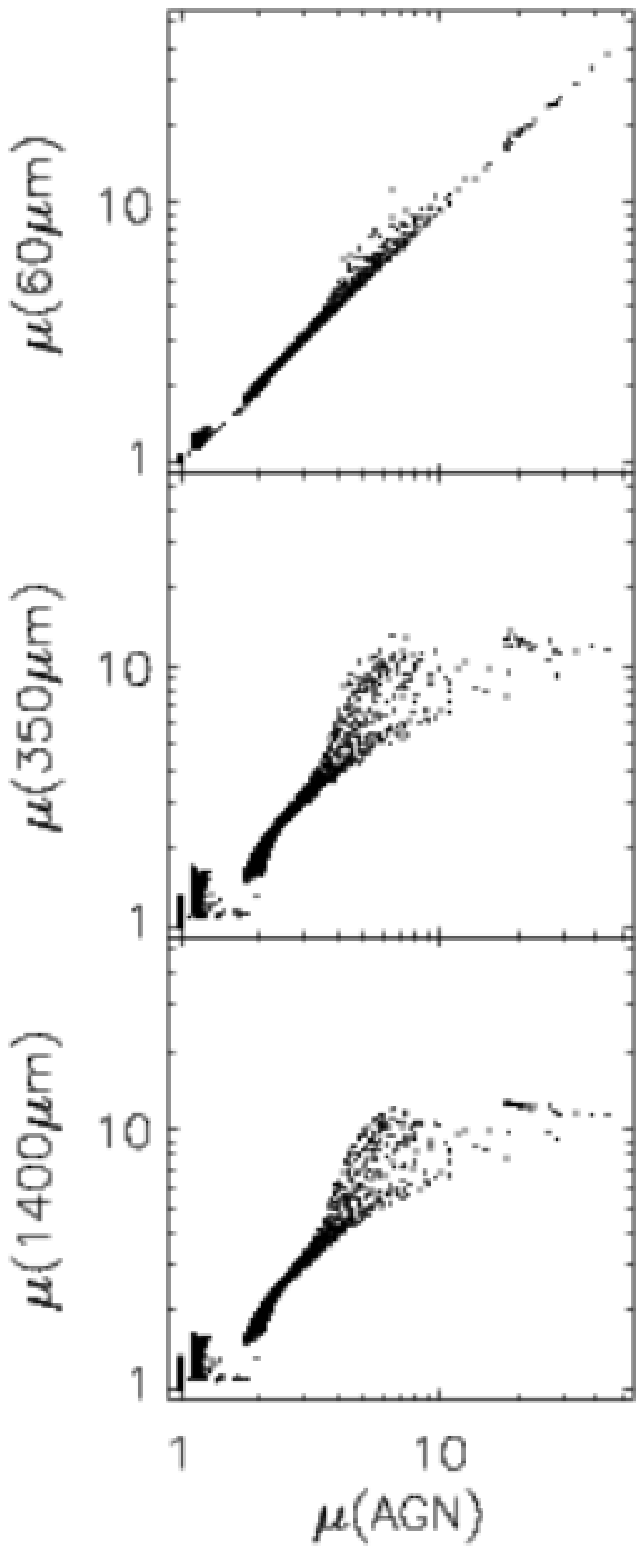}
%\ForceWidth{4.2in}
\ForceWidth{1.01in}
\hSlide{-1.7cm}
\vSlide{-0.21cm}
\BoxedEPSF{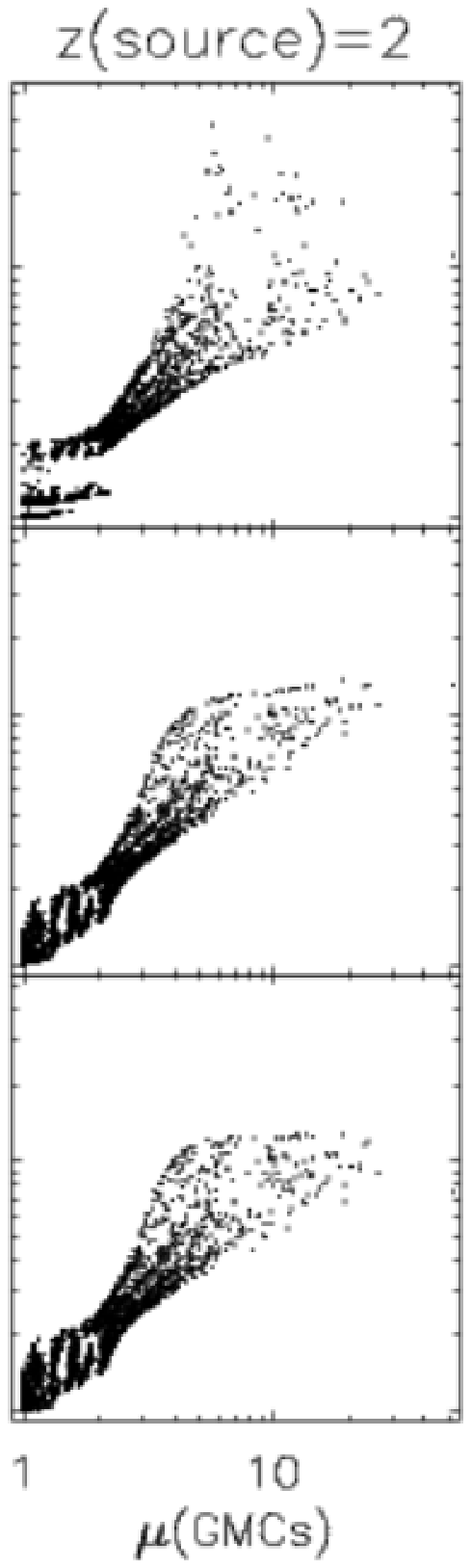}
%\ForceWidth{4.2in}
\ForceWidth{1.01in}
\hSlide{2.9cm}
\vSlide{-8.4cm}
\BoxedEPSF{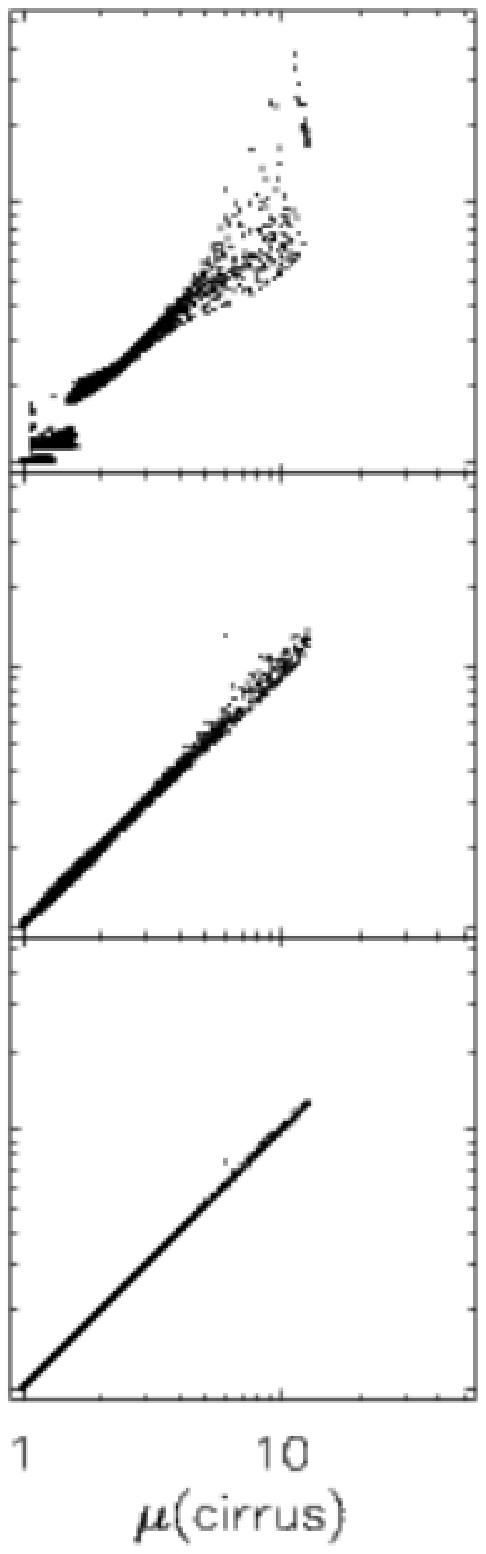}
\vspace*{-8.6cm}
\caption{\label{fig:magnification_components} The monochromatic
  magnification factors compared to the magnifications of specific
  components, for an example source at $z=2$ and a lens at $z=1$. Note
  that the AGN is well-traced by rest-frame mid-infrared photometry,
  and cirrus by photometry on the Rayleigh-Jeans tail, but that there
  is no single monochromatic tracer for star forming
  regions. Measuring this component, and decoupling it from the
  extended cirrus component, requires resolved imaging in the submillimetre or
  millimetre wavelengths.}
\end{figure}
}

}

\subsection{Source count model predictions}
One potential bias not accounted for in the discussion so far is the
effect of the slope of the luminosity function. As the slope steepens,
the magnification bias is stronger. Might one expect more pathological
lens configurations as one samples populations progressively brighter
than the break luminosity $L_*$?

\begin{figure}
\ForceWidth{4.5in}
\hSlide{-0.5cm}
\BoxedEPSF{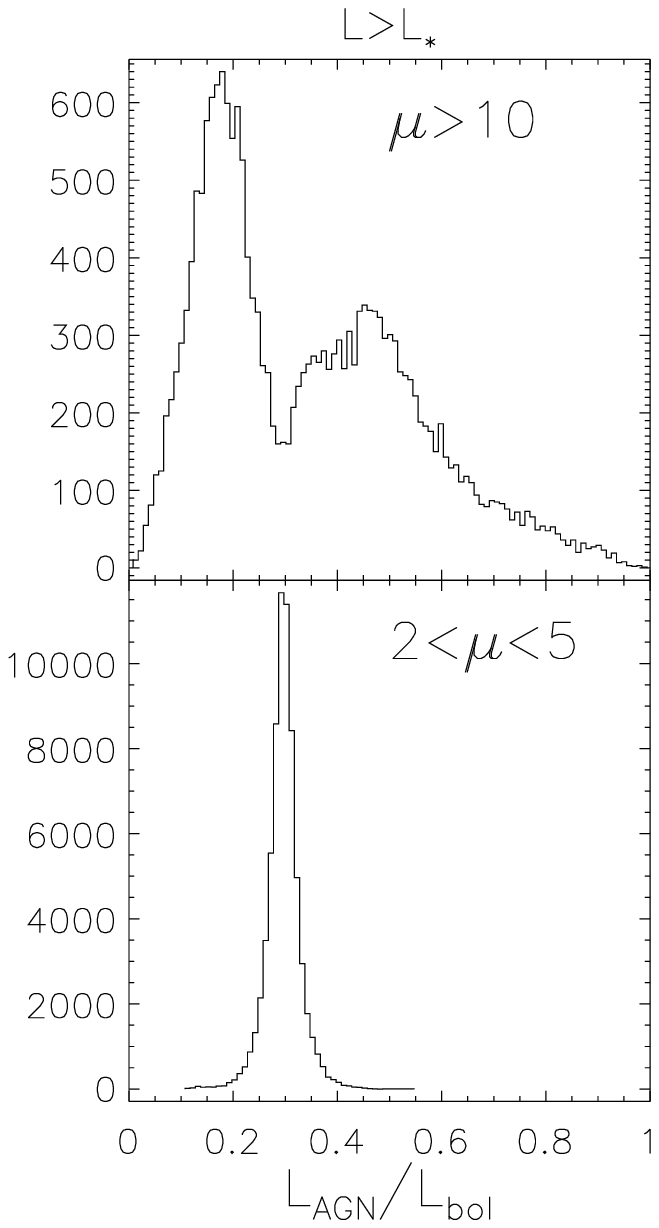}
\ForceWidth{4.5in}
\hSlide{3.1cm}
\vSlide{-8.2cm}
\BoxedEPSF{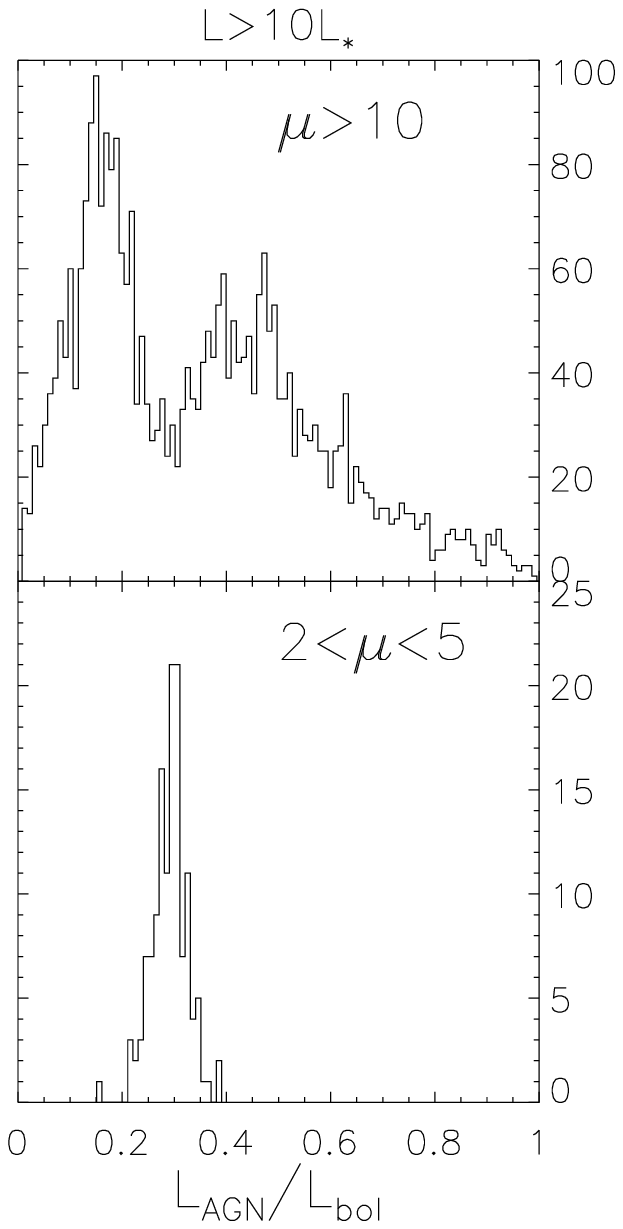}
\vspace*{-8cm}
\caption{\label{fig:schechter} Histograms of simulated AGN bolometric
  fractions for 
  $z=4$ lensed galaxies, selected at an observed wavelength of
  $500\,{\rm \mu}$m from a $100\,{\rm \mu}$m rest-frame 
  Schechter function
  with a faint-end slope of $-1.46$. 
  Samples are shown brigher than $L_*$ (left) or $10L_*$ (right). 
  The underlying bolometric
  fraction is $0.3$. The moderate magnification
  ($2<\mu<5$) and strong magnification ($\mu>10$) cases are
  shown. Qualitatively similar results are found at other source
  redshifts, and the results are only weakly dependent on the
  faint-end slope of the luminosity function.}
\end{figure}

Lapi et al. (2011) recently presented constraints on the evolving
$100\,{\rm \mu}$m rest-frame luminosity function of submillimetre-selected
galaxies. Their $z=2.4-4$ luminosity function, and their toy model
that fits it, are both consistent with a Schechter function with a
faint-end slope of $-1.46$. This luminosity function was assumed to
hold at $z=4$. Submillimetre galaxies were sampled from this distribution and
magnified by the constant comoving density of lenses described
above. The results are shown in Fig.\,\ref{fig:schechter}. It is
clear that the qualitative results of the magnification simulations in
Section \ref{sec:bolometric_fractions} are unaffected.

\section{Discussion}\label{sec:discussion}

The obvious underlying principle of these simulations is that 
{\bfreferee spatially-compact regions}
are prone to differential magnification,
particularly in the more strongly-magnified systems, while 
{\bfreferee spatially-extended regions}
are less prone to this bias. For the specific SMG
model galaxy and foreground lens population discussed in this paper,
this underlying principle leads to several specific conclusions with
clear implications for the follow-ups of infrared-selected strong
gravitational lens systems. 

It is particularly clear that for high-magnification lenses
($\mu>10$), the observed selection wavelength has a strong effect on
the distribution of relative bolometric fractions. Radiative transfer
models are an extremely effective and well-established route to
decomposing the relative bolometric fractions of ultraluminous
galaxies (e.g. Rowan-Robinson 1980; 
Rowan-Robinson \& Crawford 1989; 
Green \& Rowan-Robinson 1996; 
Granato, Danese \& Franceschini 1996; 
Silva et al. 1998, 2011; 
Efstathiou et al. 2000; 
Popescu et al. 2000, 2011;  
Misiriotis et al. 2001; 
Tuffs et al. 2004; 
M\"{o}llenhoff et al. 2006), 
but in the case of strong infrared-selected lenses with
magnifications $\mu>10$, differential magnification is so strong as to
make integrated SEDs completely unreliable in estimating relative
bolometric fractions in lensed galaxies, unless a lens model is
available to determine the differential magnification
(c.f. e.g. Granato, Danese \& Franceschini 1996, Rowan-Robinson 2000,
Farrah et al. 2002, Verma et al. 2002, 
Efstathiou 2006). Nevertheless,
for more moderate lensed systems ($2<\mu<5$), the underlying
bolometric fractions can still be reliably determined even when
neglecting the effects of differential magnification.

{\bfreferee These results are robust to the assumed configuration of the
  components in the background source. The background source would
  have to be broadly homogeneous in order to avoid strong differential
  magnification effects. One way to do this would be to argue that the
  number of GMC regions is much larger, perhaps numbering several tens
  to resemble those in the Milky Way or M31 (e.g. Sheth et
  al. 2008). However, there is little evidence that star formation in
  high-redshift infrared-luminous galaxies so closely resembles $z=0$
  spiral discs. Furthermore, an active nucleus will inevitably create
  a mid/FIR colour gradient in the background source.  }

Despite the lens magnification, the prospects are excellent for using
e.g. the [C{\sc ii}]/FIR vs. CO/FIR line diagnostic
diagram (e.g. Stacey et al. 2010, Valtchanov et al. 2011) for
determining the physical conditions in the emission line gas. The
line ratios of HCN/HNC and [O{\sc ii}]/[C{\sc ii}] appear to be
particularly insensitive to differential magnification. 
{\bfreferee The differential magnification effects on the H$_2$O lines can be
  neglected for moderate magnifications ($2<\mu<5$) but not for higher
  magnifications.}
Other
line diagnostics are nonetheless prone to differential magnification,
such as the CO ladder, even for low magnification systems
($\mu>2$). For example, the line ratio CO$(J=6-5)$/CO$(J=1-0)$ has a
$1\sigma$ dispersion of $\sim30\%$ due purely to the effects of
differential magnification for all the simulations presented in this
paper. 
{\bfreferee One might try to avoid these strong differential magnification
  effects by arguing the CO transitions are spatially co-located, but
  this is unlikely to be the case. Even in local star-forming
  galaxies, the low $J$ transitions of CO are more spatially extended
  than higher transitions (e.g. Paglione et al. 2004, Sakamoto et
  al. 2011), and evidence is accumulating of similar trends in
  high-redshift infrared-luminous galaxies (e.g. Ivison et
  al. 2011). }

There is therefore a set of photometric and spectroscopic diagnostics
that do not require a detailed lens magnification model, and others
for which such a model is beneficial or even essential. Snapshot
observations with the Atacama Large Millimetre Array ({\it ALMA}),
{\bfreferee e-Merlin, the Expanded Very Large Array (eVLA), 
the Plateau de Bure Interferometer (PdBI) and/or the 
Combined Array for Research in Millimeter-Wave Astronomy (CARMA)} 
of lens candidates from
{\it Herschel}, {\it Planck}, {\bfreferee SPT} or SCUBA-2 would easily
establish whether a system is in a low-magnification (e.g. $\mu\simeq
2$) or high-magnification (e.g. $\mu>10$) regime, {\bfreferee because
  the image separation would be enough to constrain the lens
  magnification.}  Ground-based optical or near-infrared imaging may
also assist, but the very red colours of some lensed submillimetre
galaxies (e.g. Hopwood et al. 2011) suggest that {\it HST}
near-infrared observations will often be required to detect the
background source.  {\bfreferee High-resolution
  submillimetre/millimetre-wave interferometry is also the only way to
  decouple the lensed star-forming knots from the lensed extended
  submillimetre emission (Figs.\,\ref{fig:magnification_wavelength} and
  \ref{fig:magnification_components}), giving an additional clear
  rationale for ALMA and SMA follow-ups.

  This paper has focussed on FIR to millimetre-wave emission line
  diagnostics, but similar considerations may be applicable to
  rest-frame optical and ultraviolet lines. The presence of
  Lyman\,$\alpha$ emission in some heavily-obscured
  submillimetre-selected galaxies (e.g. Chapman et al. 2005)
  necessarily requires highly anisotropic obscuration, perhaps through
  the evacuation of cavities via supernova-driven winds. Differential
  magnification may clearly affect the occurence of nebular emission
  lines in lensed infrared-luminous galaxies. If the nebular lines are
  close to or coincident with the GMC components in the galaxy, then
  there may be examples of submillimetre/millimetre-wave-selected
  lensed infrared-luminous galaxies in which the UV/optical emission
  lines are more easily detectable. A similar consideration may
  explain the strong narrow AGN emission lines in
  IRAS\,FSC\,10214$+$4724, and differential magnification may also
  help explain their unusual line ratios and widths (e.g. Serjeant et
  al. 1998). However, it does {\it not} necessarily follow that close
  proximity leads to enhancement. The model GMCs have a $50$\,pc
  radius, embedded in a warm CO knot of $400$\,pc radius, yet the
  high-$J$ CO lines do not necessarily share the total magnification
  boost of the GMCs. Similarly, extended Lyman\,$\alpha$ nebulosity
  may not necessarily be as boosted as the continuum sources at the
  selection wavelength.

  If the images of the background source 
  are resolved and a lens mass model can be
  constructed, one might alternatively choose to regard differential
  magnification as an asset, rather than a problem. Lensing conserves
  surface brightness, so high magnification regions also have high
  angular magnification. The presence of caustics in the foreground
  lens magnification could be used to make uniquely high angular
  resolution reconstructions of active nuclei or star-forming regions
  in infrared-luminous galaxies. 

\section{Conclusions}\label{sec:conclusions}

In infrared-selected gravitational lenses of any magnification
$\mu>2$, the CO ladder is strongly distorted by differential
magnification. The bolometric fractions of the FIR [C{\sc
  ii}] and [O{\sc i}] lines are also affected by differential
magnification, but the distortion is not typically sufficient to alter
the physical interpretation. Similarly, differential magnification
should not typically move a galaxy more than a factor of about two
from its position on the proposed galaxy ``main sequence''. For
moderate-magnification lenses ($2<\mu<5$), the FIR H$_2$O,
HCN and HNC lines are broadly unaffected by differential
magnification, and the broadband SED-based decompositions of
bolometric contributions should be reliable. However, for strong
lenses ($\mu>10$) 
in the absence of a foreground mass model, 
the H$_2$O lines are significantly distorted, and
unless a detailed lens model can be constructed, 
the apparent bolometric contributions of AGN, GMCs and cirrus derived
from the SEDs are so unreliable as to be useless.  }

\section{Note added in proof}
After this paper was accepted, Hezaweh et al. (2012) presented a
complementary study of differential magnification for a source
population with varying physical sizes. They found that
highly-magnified populations over-represent compact physical
components in good agreement with Section 3.4 of this paper. 

\section{Acknowledgements}
This research was supported by STFC under grant ST/G002533/1. The
author would like to thank 
{\bfreferee Shane Bussmann, }
Loretta Dunne, Rob Ivison, Roxana Lupu, Mattia
Negrello, Douglas Scott, Ian Smail 
{\bfreferee and Glenn White}
for stimulating discussions,
{\bfreferee and the anonymous referee for very helpful and timely comments.}


\begin{thebibliography}{}

\bibitem{} Aalto, A., et al., 2009, A\&A, 493, 481
\bibitem{} Auger, M.W., et al., 2009, ApJ, 705, 1099
\bibitem{} Bayliss, M.B., et al., 2011, ApJ, 727, L26
\bibitem{} Blain, A., 1996, MNRAS, 283, 1340
\bibitem{} Blain, A., 1999, MNRAS, 304, 669
\bibitem{} Bolton, A.S., et al., 2006, ApJ, 638, 703
\bibitem{} Bolton, A.S., et al., 2008, ApJ, 682, 964
\bibitem{} Broadhurst, T., Leh\'ar, J., 1995, ApJ 450, L41
\bibitem{} Brown, R.L., Vanden Bout, P.A., 1991, AJ, 102, 1956
\bibitem{} Bullock, J.S., et al., 2001, MNRAS, 321, 559
\bibitem{} Carilli, C.L., et al., 2010, ApJ, 714, 1407
\bibitem{} Chapman, S.C., et al., 2005, ApJ, 622, 772
\bibitem{} Contursi, A., et al., 2010, proceedings of the Herschel
  First Results Symposium (ESLAB 2010)
\bibitem{} Cooray, A., et al., 2010, arXiv:1007.3519
\bibitem{} Cooray, A., et al., 2011, ApJ, 732, L35
\bibitem{} Cox, P., et al., 2011, ApJ, 740, 63
\bibitem{} Daddi, E., et al., 2007, ApJ, 670, 156
\bibitem{} Deane, R.P., Rawlings, S., Marshall, P., Heywood, I.,
  Kl\"{o}ckner, H.-R., Grainge, K., Mauch, T., Serjeant, S., 2011,
  MNRAS, submitted
\bibitem{} Eales, S.A., et al., 2010, PASP, 122, 499
\bibitem{} Eisenhardt, P., Armus, L., Hogg, D.W., Soifer, B.T.,
Neugebauer, G., Werner, M.W., 1996, ApJ 461, 72
\bibitem{} Efstathiou, A., 2006, MNRAS, 371, L70
\bibitem{} Efstathiou, A., Rowan-Robinson, M., \& Siebenmorgen, R., 2000, MNRAS, 313, 734
\bibitem{} Elbaz, D., et al., 2007, A\&A, 468, 33
\bibitem{} Farrah, D., Serjeant, S., Efstathiou, A., Rowan-Robinson,
  M., Verma, A., 2002, MNRAS, 335, 1163
\bibitem{} Frayer, D.T., et al., 2011, ApJ, 726, L22
\bibitem{} Gavazzi, R., et al., 2007, ApJ, 667, 176
\bibitem{} Gavazzi, R., et al., 2011, ApJ, 738, 125
\bibitem{} Gonz\'alez, V., et al., 2010, ApJ, 713, 115
\bibitem{} Gonz\'alez-Alfonso, E., et al., 2010, A\&A, 518, L43
\bibitem{} Gonz\'alez-Nuevo, J., et al., 2012, ApJ, 749, 65
\bibitem{} Graham, J.R., Liu, M.C., 1995, ApJ, 449, L29
\bibitem{} Granato, G.L., Danese, L., Franceschini, A., 1996, ApJ,
  460, L11
%\bibitem{} Granato, G.L., et al., 2001, MNRAS, 324, 757
\bibitem{} Green, S.M., \& Rowan-Robinson, M., 1996, MNRAS, 279, 884
\bibitem{} Hainline, L., et al., 2011, ApJ, 740, 96
\bibitem{} Hailey-Dunsheath, S., et al., 2010, ApJ, 714, L162
\bibitem{} Hezaweh, Y.D., Marrone, D.P., Holder, G.P., 2012, preprint
  (arXiv:1203.3267) 
\bibitem{} Hoekstra, H., Hseih, B.C., Yee, H.K.C., Lin, H., 
Gladders, M.D., 2005, ApJ, 635, 73
\bibitem{} Hopwood, R., et al., 2011, ApJ, 728, L4
\bibitem{} Ivison, R.J., et al., 2011, MNRAS, 412, 1913
\bibitem{} Keeton, C.R., 2001, preprint (astro-ph/0102340)
\bibitem{} Keeton, C.R., Kochanek, C.S, 1998, ApJ, 495, 157
%\bibitem{} Knudsen, K.K., et al., 2007, ApJ, 666, 156  %% NGC 253
\bibitem{} Knudsen, K.K., et al., 2008, MNRAS, 384, 1161
%\bibitem{} Kuno, N., et al., 2007, PASJ, 59, 117 %% NGC 253
\bibitem{} Lacy, M., Rawlings, S., Serjeant, S., 1998, MNRAS, 299, 1220
\bibitem{} Lapi, A., et al., 2011, ApJ, 742, 24
\bibitem{} Loenen, A.F., Boan, W.A., Spaans, M., 2007, proceedings of IAU
  Symposium 242, Astrophysical Masers and their Environments,
  eds. J.M. Chapman and W.A. Baan, CUP, Cambridge, pp 462-466 
  (arXiv:0709.3423) 
\bibitem{} Lupu, R., et al., 2010, ApJ submitted (arXiv:1009.5983)
\bibitem{} Lupu, R., et al., 2011, in The Molecular Universe,
  Proceedings of the 280th Symposium of the International Astronomical
  Union, eds. Jos\'{e} Cerricharo and Rafael Bachiller, Toledo, Spain,
  May 30-June 3 2011, CUP Cambridge, in press.
\bibitem{} McKee, C.F., Ostriker, E.C., 2007, ARA\&A, 45, 565
\bibitem{} Micha{\l}owski, M., et al., 2012, A\&A, 541, 85
\bibitem{} Misiriotis, A., Popescu, C.C., Tuffs, R., Kylafis, N.D.,
  2001, A\&A, 372, 775
\bibitem{} M\"{o}llenhoff, C., Popescu, C.C., Tuffs, R.J., 2006, A\&A,
  456, 941
\bibitem{} M\"{o}ller, O., Kitzbichler, M., Natarajan, P., 2007,
  MNRAS, 379, 1195
\bibitem{} Myers, S.T., et al., 2003, MNRAS, 341, 1
\bibitem{} Negrello, M., et al., 2007, MNRAS, 377, 1557
\bibitem{} Negrello, M., et al., 2010, Science, 330, 800
\bibitem{} Nenkova, M., Sirocky, M.M., Ivesi$\acute{\rm c}$,
  $\breve{\rm Z}$., Elitzur, M, 2008a,
  ApJ, 685, 147
\bibitem{} Nenkova, M., Sirocky, M.M., Ivesi$\acute{\rm c}$,
  $\breve{\rm Z}$., Elitzur, M, 2008b,
  ApJ, 685, 160
\bibitem{} Noeske, K.G., et al., 2007, ApJ, 660, L43
\bibitem[Omont et al. 2011]{Omont11}
{Omont, A., et al.}, 2011, \textit{A\&A}, 530, L3
%\bibitem{} Paglione, T.A.D., et al., 2001, ApJS, 135, 183 %% NGC 253
\bibitem{} Paglione, T.A.D., et al., 2004, ApJ, 611, 835 %% NGC 253
\bibitem{} Perrotta, F., Baccigalupi, C., Bartelmann, M., De Zotti,
  G., Granato, G.L., 2002, MNRAS, 339, 445
\bibitem{} Polletta, M., et al. 2007, ApJ, 663, 81
\bibitem{} Popescu, C.C., Misiriotis, A., Kylafis, N.D., Tuffs, R.J.,
  Fischera, J., 2000, A\&A, 362, 138
\bibitem{} Popescu, C.C., Tuffs, R.J., Dopita, M.A., Fischera, J.,
  Kylafis, N.D., Madore, B.F., 2011, A\&A, 527, 109
\bibitem{} Riechers, D.A., et al., 2010, ApJ, 720, L131
\bibitem{} Riechers, D.A., et al., 2011, ApJ, 733, L12
\bibitem{} Rowan-Robinson, M., 1980, ApJS, 44, 403
\bibitem{} Rowan-Robinson, M., 2000, MNRAS, 316, 885
\bibitem{} Rowan-Robinson, M., Crawford, J., 1989, MNRAS, 238, 523
\bibitem{} Rowan-Robinson, M., et al., 1991, Nature 351, 719
%\bibitem{} Sakamoto, K., et al., 2006, ApJ, 636, 685 %% NGC 253
\bibitem{} Sakamoto, K., et al., 2011, ApJ, 735, 19 %% NGC 253
\bibitem{} Scott, K.S., et al., 2011, ApJ, 733, 29
\bibitem{} Serjeant, S., 2010, {\it Observational Cosmology},
  Cambridge University Press.
\bibitem{} Serjeant, S., 2011, in Proceedings of IAU Symposium 284,
  The Spectral Energy Distribution of Galaxies, 
  eds. Richard J. Tuffs and Cristina C. Popescu, CUP, Cambridge, in
  press (arXiv:1112.0323)
\bibitem{} Serjeant, S., Lacy, M., Rawlings, S., King, L.J., Clements,
D.L., 1995, MNRAS, 276, L31
\bibitem{} Serjeant, S., et al., 1998, MNRAS, 298, 321
\bibitem{} Sheth, K., et al., 2008, ApJ, 675, 330
\bibitem{} Silva, L., Granato, G.L.; Bressan, A., Danese, L., 1998,
  ApJ, 509, 103
\bibitem{} Silva, L., et al., 2011, MNRAS, 410, 2043
\bibitem{} Smail, I., et al., 1997, ApJ, 490, L5
%\bibitem{} Sorai, K., et al., 2000, PASJ, 52, 785 %% NGC 253
\bibitem{} Stacey, G., et al., 2010, ApJ, 724, 957
\bibitem{} Stark, D.P., et al., 2009, ApJ, 697, 1493
\bibitem{} Sturm, E., et al., 2010, ApJ, 733, L15
\bibitem{} Swinbank, A.M., et al., 2010, Nature, 464, 733
\bibitem{} Tacconi, L.J., et al., 2006, ApJ, 640, 228
\bibitem{} Treu, T., 2010, ARA\&A, 48, 87
\bibitem{} Tuffs, R.J., Popescu, C.C., V\"{o}lk, H.J., Kylafis, N.D.,
  Dopita, M.A., 2004, A\&A, 419, 821
\bibitem{} Walter, F., et al., 2009, Nature, 457, 699
\bibitem{} Wei\ss, A., Henkel, C., Downes, D., Walter, F., 2005, A\&A,
  409, L41
\bibitem{} Valtchanov, I., et al., 2011, MNRAS, 415, 3473
\bibitem{} van der Tak, F.F.S., et al., 2007, A\&A, 468, 627
\bibitem{} van der Werf, P.P., et al., 2010, A\&A, 518, L42
\bibitem{} van der Werf, P.P., et al,. 2011, ApJ, 741, L38
\bibitem{} Verma, A., Rowan-Robinson, M., McMachon, R., Efstathiou,
  A., 2002, MNRAS, 335, 574
\bibitem{} Vieira, J.D., et al., 2010, ApJ, 719, 763
\bibitem{} Wuyts, E., et al., 2012, ApJ, 745, 86

\end{thebibliography}
\end{document}